\documentclass[12pt]{article}
\usepackage{fullpage,amsmath}
\usepackage{setspace}
\usepackage{graphicx, psfrag, amsfonts, bm}
\usepackage{natbib}
\usepackage{wrapfig}
\usepackage{multirow}
\usepackage{hhline}
\usepackage{caption}
\usepackage{url}

\def\btheta{\bm{\theta}}

\def\bsig{\bm{\sigma}}
\def\bphi{\bm{\phi}}

\def\bpi{\bm{\pi}}
\def\bpit{\bm{\widetilde{\pi}}}
\def\pit{\widetilde{\pi}}

\newcommand{\bx}{\bm{x}}

\newcommand{\bn}{\bm{n}}
\newcommand{\bN}{\bm{N}}

\newcommand{\bZ}{\bm{Z}}
\newcommand{\bz}{\bm{z}}
\newcommand{\bL}{\bm{L}}
\newcommand{\bell}{\bm{\ell}}
\newcommand{\bw}{\bm{w}}

\newcommand{\ba}{\bm{a}}

\newcommand{\Cstar}{C^\star}
\newcommand{\Cs}{C^\star}
\newcommand{\true}{{\mbox{\tiny TRUE}}}   
\renewcommand{\th}{\theta}
\newcommand{\sig}{\sigma}

\newcommand{\bth}{\bm{\th}}

\newcommand{\Ct}{\widetilde{C}}

\newcommand{\bxt}{\widetilde{\bx}}

\newcommand{\DD}{{\cal D}}

\newcommand{\DU}{\mbox{DU}}

\newcommand{\Ga}{\mbox{Gamma}}
\newcommand{\Dir}{\mbox{Dir}}

\newcommand{\Be}{\mbox{Be}}

\newcommand{\Binom}{\mbox{Bin}}
\newcommand{\Geom}{\mbox{Geom}}
\newcommand{\Poi}{\mbox{Poi}}
\newcommand{\BeDir}{\mbox{Be-Dir}}
\newcommand{\iid}{\stackrel{iid}{\sim}}
\newcommand{\indep}{\stackrel{indep}{\sim}}

\usepackage[dvipsnames,usenames]{color}
\newcommand{\bch}{\color{blue}\it}
\newcommand{\ech}{\color{Black}\rm}

\newcommand{\note}[1]{\color{blue} {\sc [Note]}\footnote{\color{Brown}\rm #1
    \color{Black}} \color{Black}}

\begin{document}
\doublespacing

\title{Bayesian Inference for Tumor Subclones Accounting for Sequencing and Structural Variants}

\author{
        Juhee Lee, \\
        {\small Department of Applied Mathematics and Statistics, University of California Santa Cruz}        
		\and        
        Peter M\"uller%\thanks{Correspondence:  Department of Mathematics UT Austin 1, University Station, C1200 Austin, TX 78712 USA. E-mail: pmueller@math.utexas.edu}
        ,\\
        {\small Department of Mathematics, University of Texas Austin}
        \and 
         Subhajit Sengupta, \\
         {\small Center for Clinical Research and Informatics,
           Northshore University HealthSystem}
        \and
        Kamalakar Gulukota \\         
        {\small Center for Molecular Medicine, Northshore University HealthSystem}       
         \and
         Yuan Ji\thanks{Correspondence: 1001 University Place,
           Evanston, IL 60201. E-mail: jiyuan@uchciago.edu}\\
         {\small Center for Clinical Research and Informatics,
           Northshore University HealthSystem} \\ 
           {\small Department of Health
           Studies, The University of Chicago}       
}

\date{}
\maketitle

\begin{abstract}
\noindent
Tumor samples are heterogeneous.  They consist of different
subclones
 that are characterized by differences in 
DNA nucleotide sequences and copy numbers on multiple loci.  
Heterogeneity can be measured through the identification of the
subclonal copy number and sequence at  a selected set of loci. 
Understanding that the accurate identification of variant
allele fractions greatly depends on a precise determination of copy
numbers, we develop a Bayesian feature allocation model for jointly
calling subclonal copy numbers and the corresponding allele sequences
for the same loci. The proposed method utilizes three random matrices,
$\bL$, $\bZ$ and $\bw$ to represent subclonal copy numbers ($\bL$),
 numbers of subclonal  variant alleles ($\bZ$) and cellular
fractions of subclones in samples ($\bw$), respectively.  
 The unknown number of subclones implies
a random number of columns for these matrices. 
 We use  next-generation sequencing data
to estimate the subclonal structures through 
 inference on these three matrices.
 Using  simulation studies and a real data analysis, 
 we demonstrate how 
posterior inference on the subclonal structure is enhanced  with
 the
joint modeling of both structure and sequencing variants on subclonal
genomes.   %The proposed inference provides a complete depiction of subclonal DNA.   
Software is available at \url{http://compgenome.org/BayClone2}.

\end{abstract}

%%%%%%%%%%%%%%%%%%%%%%%%%%%%%%%%%%%%%%%%%%%%%%%%%%%%%%%%%%%%%%%%%%%%%%%%%%%%%%
\section{Introduction}
\label{sec:Intro_intra}
%%%%%%%%%%%%%%%%%%%%%%%%%%%%%%%%%%%%%%%%%%%%%%%%%%%%%%%%%%%%%%%%%%%%%%%%%%%%%%
\subsection{Biological background and motivation}
Understanding tumor heterogeneity (TH) is critical for
precise cancer prognosis. Not 
all tumor cells have the same genome and respond to the same
treatment. 
TH arises when somatic mutations occur in only a 
fraction of tumor cells, and results in the 
observed spatial and temporal heterogeneity of tumor samples
\citep{Russnes:hetero:2011, greaves2012clonal, frank2004problems,
biesecker2013genomic, frank2003cell, de2011somatic, bedard2013tumour,
navin2011tumour, ding2012clonal}.  
 In other words, a tumor sample is composed of different subclones of cells
 with each subclone being defined by a
unique genome.
Figure \ref{fig:data_1}(a) illustrates this process  with a
hypothetical case in which accumulation of variants over the lifetime
of a tumor gives rise to different subpopulations of tumor cells.
Researchers have
recently started to recognize the importance of TH and realize the
mistake of treating cancer using a “one-size-fits-all”
approach. Instead, precision medicine now aims to focus on 
%\note{if they were already successfully ``aiming'', then no need for us?}
targeted
treatment of individual tumors based on their molecular
characteristics, including TH. 

Rapid progress has been
made in the development of 
computational tools for clonal inference in the past year
\citep{oesper2013theta,miller2014sciclone,Strino01092013,24484323,zare2014inferring}.
New methods continue to set new and higher standards in the
statistical inference for TH that mimic the underlying biology ever
more closely.
However, the current literature still lacks  effective methods,
computational or experimental, for assessing differences between
subclonal genomes in terms of both structure variants, such as copy
number variants (CNVs), and in terms of sequence variants,
such as single nucleotide variants (SNVs). 
More importantly, current methods lack 
computational models that could jointly estimate copy
numbers and variant allele counts within each subclone.  Recent work by
\cite{Li-GB:2014} adjusts the estimation of subclonal cellular
fractions for both CNVs and SNVs, but still stops short
of directly inferring subclonal copy numbers or variant allele counts.

% These two quatities, combined with
% the cellular fractions of all the subclones, will decide the observed
% number ($N$) of short reads mapped to a locus and the number $n (\le N)$ of short reads bearing a variant
% sequence. See Figure 1(b) as an illustration. 

%\cut %Figure \ref{fig:data_1}(a) shows a hypothetical history of tumor evolution. 
%\note{We already said what figure 1a is above, I think.}
Figure \ref{fig:data_1}(b) shows a stylized example of
%corresponding observed  
DNA-Seq data for a % heterogeneous tumor 
sample taken on day 360 of the process shown in Figure~\ref{fig:data_1}(a).
The sample is a result of the underlying
tumor evolution. The sample has
three tumor subclones.  If the sample is sequenced and short reads
are mapped, the total number of reads mapped to each locus will be
affected by the copy numbers of all the subclones. In Figure
\ref{fig:data_1}(b), due to the copy number gains 
in subclones 2 and
3, we expect that there will be additional reads with sequence A at
locus 1 and additional reads with sequence G at locus 3 (both marked
by brown letters). In addition, 
\begin{minipage}[l]{0.59\textwidth}
  % \begin{figure}
  \hskip -.2in
  \begin{tabular}{c}
    \includegraphics[width=1\textwidth]{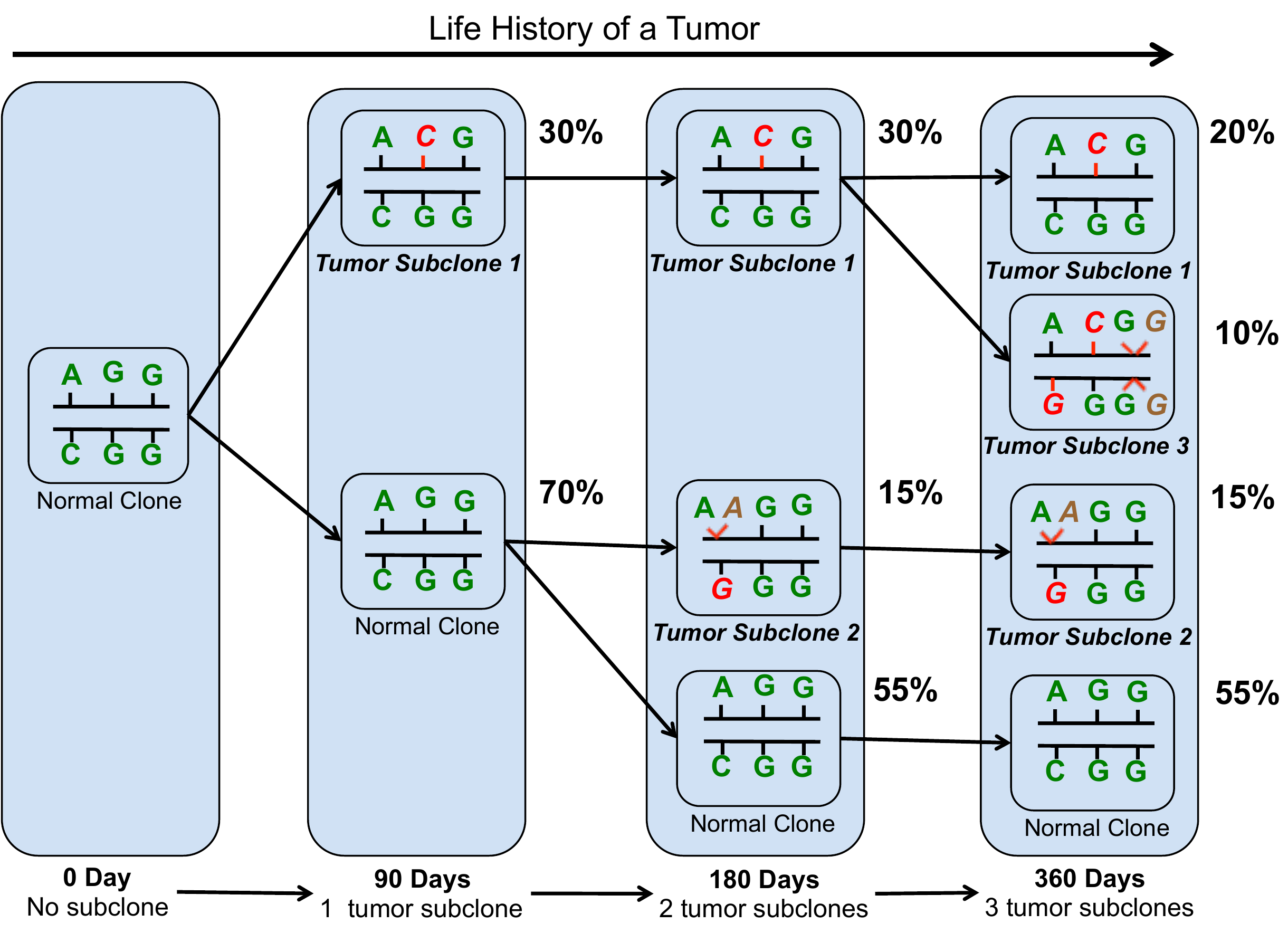} \\
    {\bf (a)}\\ %\hline
    \\
    \includegraphics[width=1\textwidth]{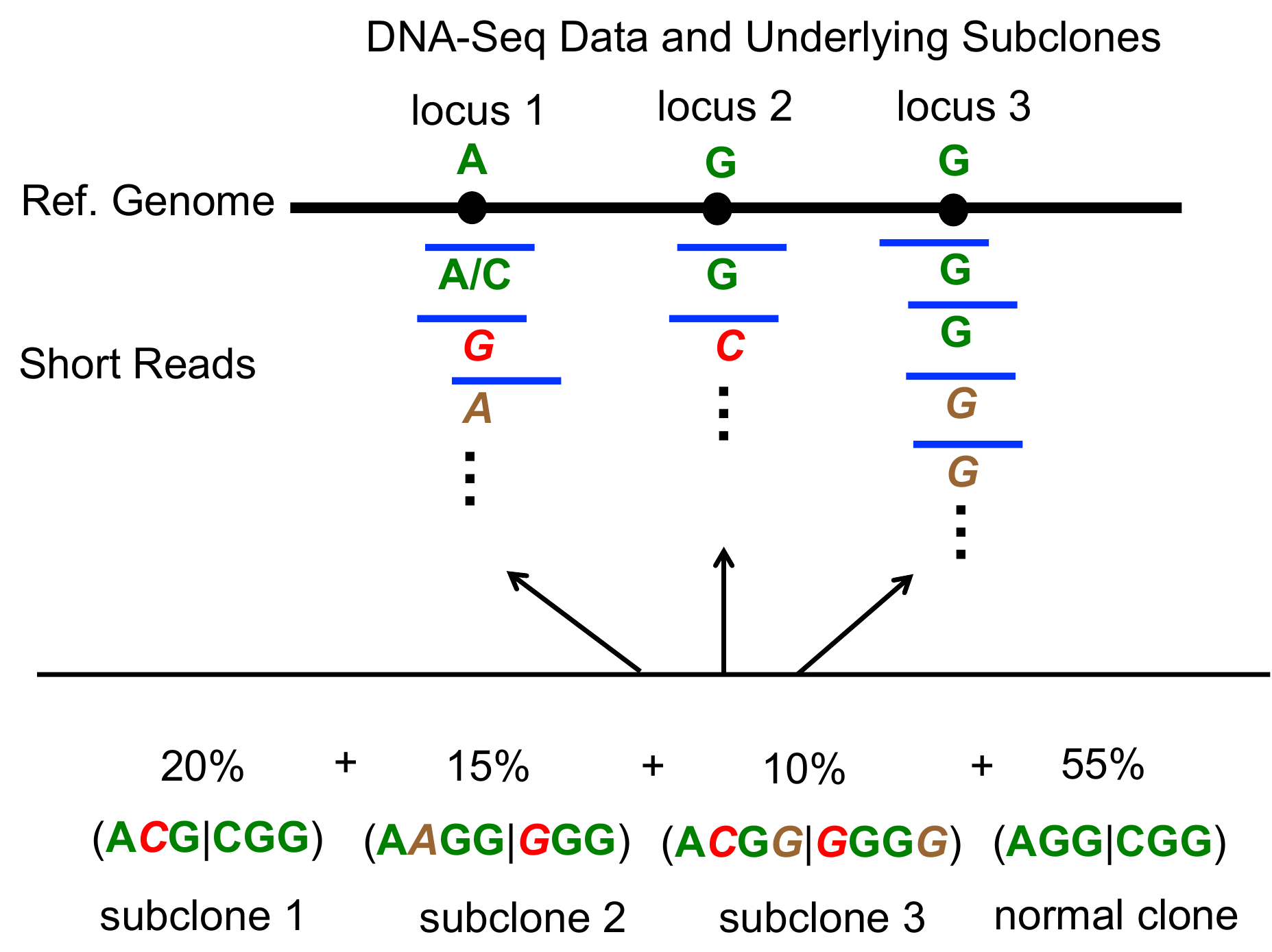} \\
    {\bf (b)}\\ %\hline
  \end{tabular}
  \hskip -.5in
  \captionof{figure}{\footnotesize {\bf (a)} Tumor heterogeneity caused by clonal expansion. On days 90, 180, and 360, four somatic mutations (represented by red letters) and three somatic copy number gains (represented by brown letters) result in three tumor subclones. {\bf (b)} Observed short reads (some with variants) are results of heterogeneous subclonal genomes. In particular, the formula at the bottom shows that subclonal alleles are mixed in proportions to produce short reads, which are mapped to different loci. %Accounting for noise, we aim to use the observed short read counts  to recover the subclonal genetic structure. 
  }
  \label{fig:data_1}
  % \end{figure}
\end{minipage}
\hskip .1in
\begin{minipage}[r]{0.35\textwidth}
  % \begin{figure}
  \vskip .7in
  \begin{tabular}{c}
    \includegraphics[page=3,height=1.691in,width=.9\textwidth]{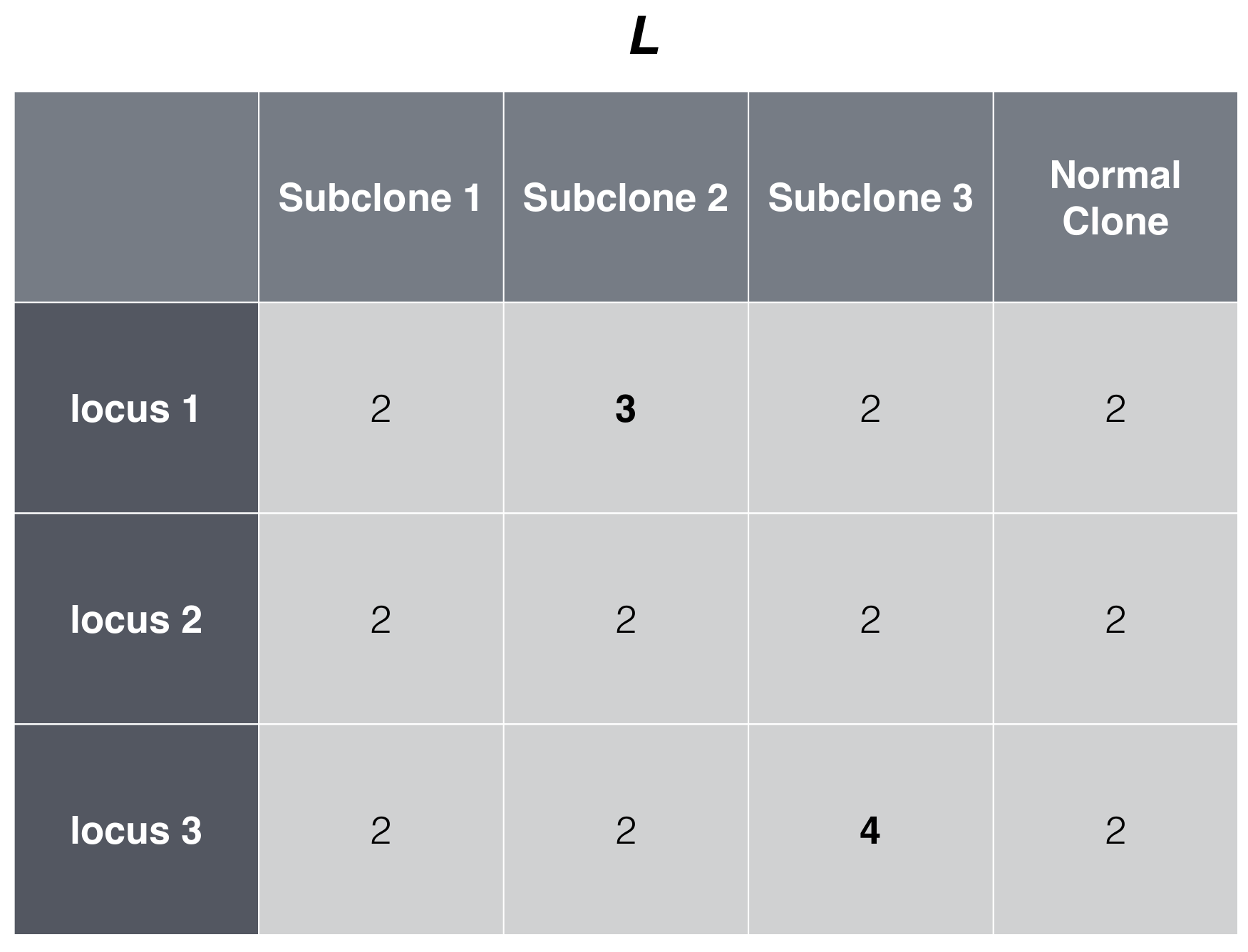} \\
    \includegraphics[page=4,height=1.691in,width=.9\textwidth]{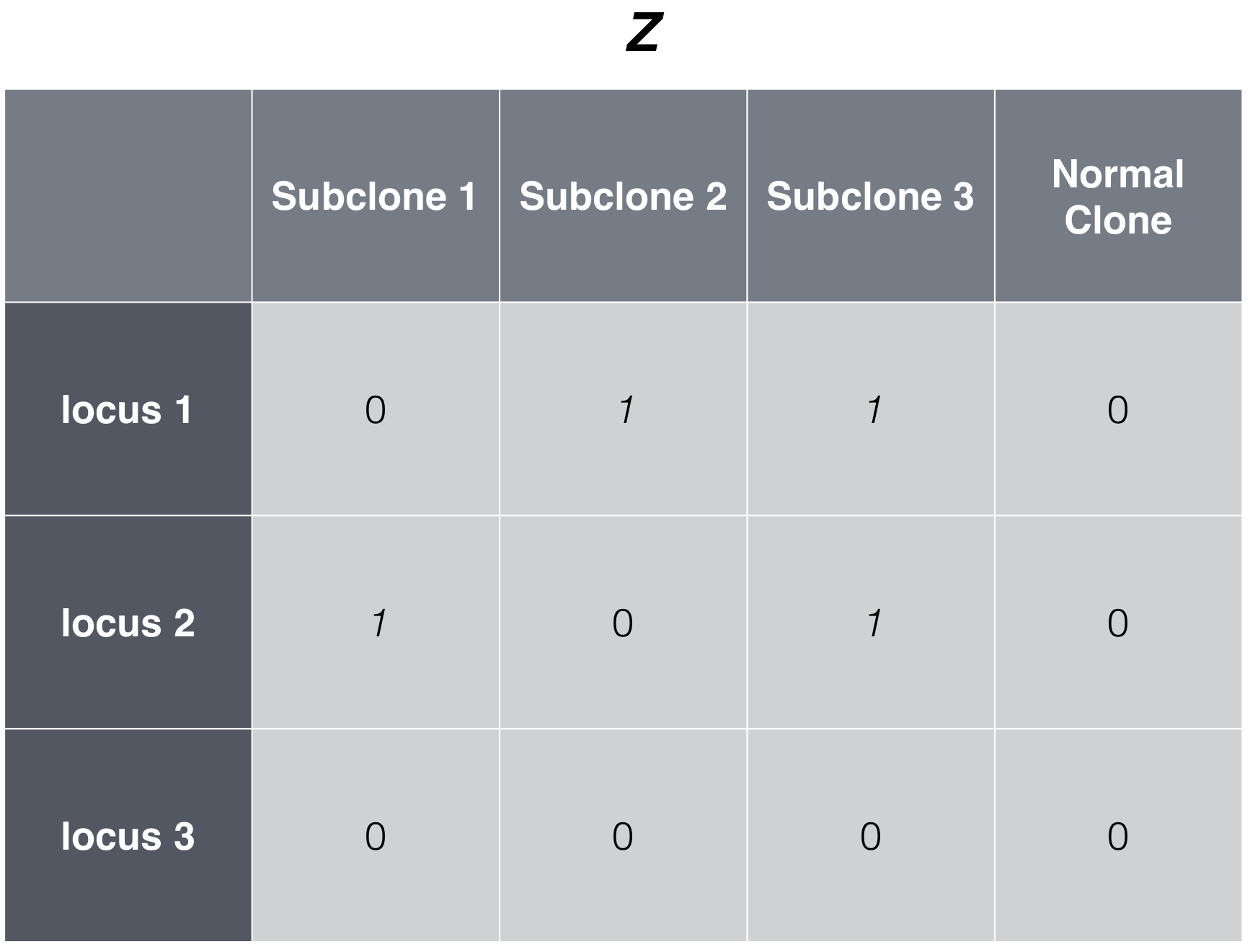} \\
    \includegraphics[page=5,height=1.691in,width=.9\textwidth]{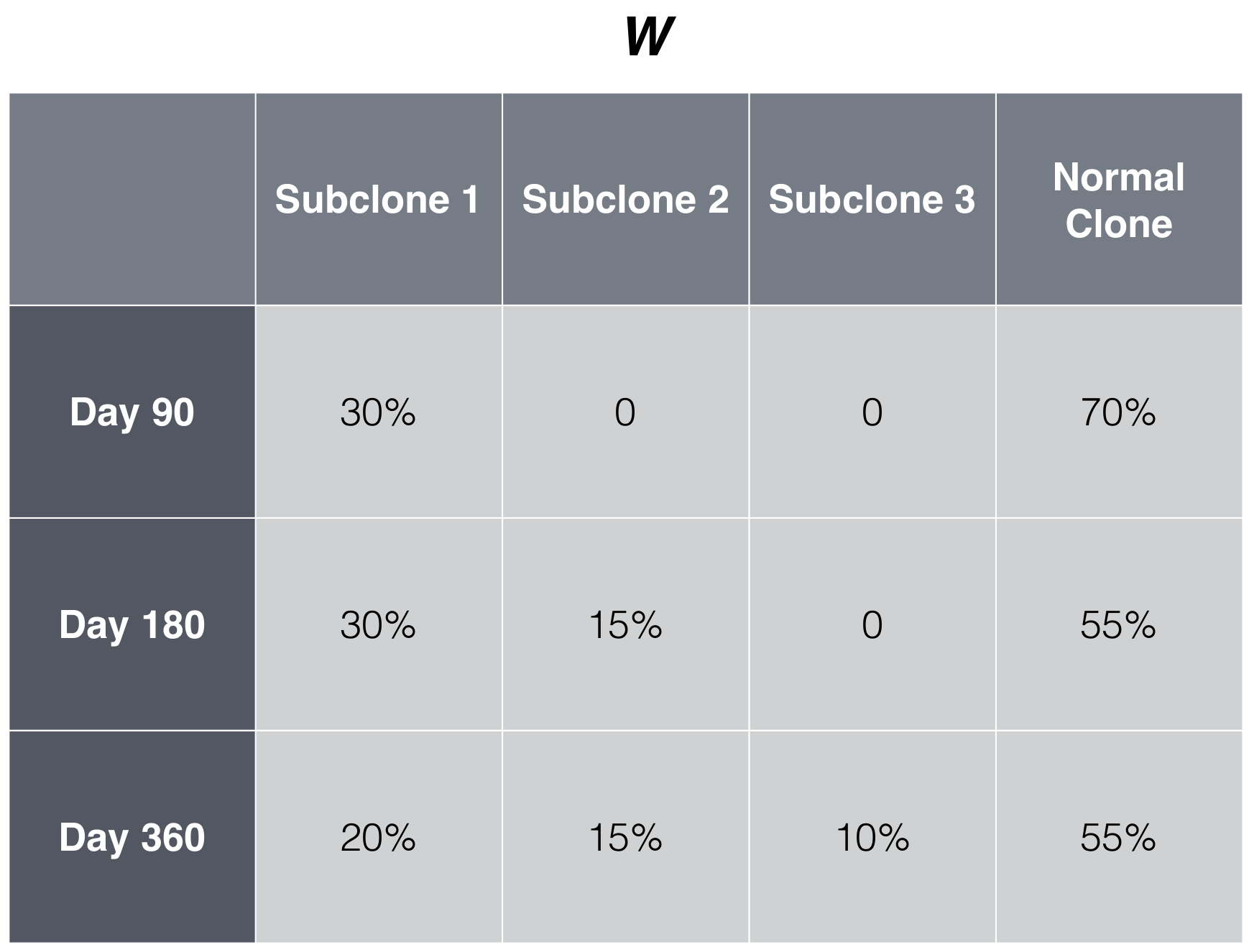} \\
  \end{tabular}
  \captionof{figure}{\footnotesize Three matrices for inference to describe the subclonal structure in Figure \ref{fig:data_1}. $\bL$ describes the subclonal copy numbers, $\bZ$ describes the numbers of subclonal variant alleles, and $\bm{w}$ describes the cellular fractions of subclones. } \label{fig:infer-target}
  \vspace{-.0in}
  % \end{figure}
\end{minipage} 
\clearpage
\noindent
variant short reads will be generated
due to the subclonal mutations in the sample, such as short reads with
the red letters mapped to loci 1 and 2.  
 Using NGS data
% Accounting for noises and artifacts in the NGS data, 
we aim to recover the subclonal sequences
at these loci and cellular fractions at the bottom of Figure
\ref{fig:data_1}(b) that explains the true biology in (a). In
particular, we aim to provide three matrices as shown in Figure
\ref{fig:infer-target} to describe the subclonal genomes and sample
heterogeneity. For illustration, 
Figure \ref{fig:infer-target} fills in the (biological) truth for these three
matrices corresponding to the hypothetical tumor heterogeneity
described in Figure \ref{fig:data_1}(a). 
In an actual data analysis, all three matrices are latent and must be estimated.

\subsection{Model-based Inference for Tumor Heterogeneity}
We propose a new class of Bayesian feature allocation models
\citep{broderick2013cluster} to  implement 
inference on these three matrices. 
We first construct an integer-valued 
matrix  $\bL$ to characterize subclonal copy numbers.  
Each column corresponds to a subclone and rows correspond to loci. We
use the column vector $\bell_c =(\ell_{1c}, \ldots, \ell_{Sc})$ of
integers to  represent copy numbers  across  $S$ loci 
for subclone $c$.  For example, in Figure~\ref{fig:infer-target},
$\ell_{sc} =3$ for $s=1$ and $c=2$ since subclone 2 has three alleles
at locus 1.  As a prior distribution for $\bL$, $p(\bL)$, we will define a finite version of a
categorical Indian buffet process
\citep{Subhajit:2013,Sengupta:2014}, a new feature allocation model.  

%\note{Sengupta S, Gulukota K, Lee J, Mueller, P, Ji Y*.   BayClone: Bayesian Nonparametric Inference of Tumor Subclones Using NGS Data. Submitted.}.  
% Similar to the construction of the Indian buffet process,  of the random integer matrix is the categorical Indian buffet process (cIBP) developed in \cite{Subhajit:2013,Sengupta:2014} \note{as the number of subclones goes to infinity.

Next, we introduce a second integer-valued matrix $\bZ$ with
the same dimensions as $\bL$.  We use $\bZ$ to record SNV's.
Denote by $\bz_c$ the $c$-th column of $\bZ$.
Conditional on $\bell_c$, $\bz_c = (z_{1c},
\ldots, z_{Sc})$, $z_{sc} \leq \ell_{sc}$,
represents the number of alleles that bear a mutant sequence
different from the reference sequence at locus $s$, $s=1, \ldots, S$
in subclone $c$.
For example, in Figure~\ref{fig:infer-target}, $z_{sc}=1$ for
$s=2$ and $c=1$, indicating that one allele bears a variant
sequence. By definition, the number of variant alleles $z_{sc}$ in a
subclone cannot be larger than the copy number $\ell_{sc}$ of the
subclone, i.e., $z_{sc} \le \ell_{sc}.$  Jointly, the two random integer vectors, $\bell_c$ and $\bz_c$
describe a subclone and its genetic architecture at the corresponding
loci.  Lastly, we introduce the $\bw$ matrix.
Each row $\bw_t=(w_{t1},\ldots,w_{tC})$ 
represents the cellular fractions of the $C$ subclones in each
sample (and we will still add an additional subclone $c=0$).

The remainder of the paper is organized as follows:
Section~\ref{sec:Model_Sel} describes the proposed Bayesian feature
allocation model. % \cut % and a model selection criterion to select the number of subclones. 
Section \ref{sec:Simulation} describes simulation
studies. Section~\ref{sec:lung_cancer} reports a data analysis for an
in-house data set to illustrate intra-tumor heterogeneity. The last
section concludes with a final discussion. 

%%%%%%%%%%%%%%%%%%%%%%%%%%%%%%%%%%%%%%%%%%%%%%%%%%%%%%%%%%%%%%%%%%%%%%%%%%%%
\section{Probability Model}
\label{sec:Model_Sel}
%%%%%%%%%%%%%%%%%%%%%%%%%%%%%%%%%%%%%%%%%%%%%%%%%%%%%%%%%%%%%%%%%%%%%%%%%%%%
\subsection{Sampling model}
\label{sec:Models}
%%%%%%%%%%%%%%%%%%%%%%%%%%%%%%%%%%%%%%%%%%%%%%%%%%%%%%%%%%%%%%%%%%%%%%%%%%%%
  Suppose that $T$ samples have been sequenced in an NGS
experiment. These samples are assumed to be from the same patient,
obtained either at different time points or different geographical
locations within the tumor.    Suppose that we have collected   read
mapping data 
on $S$ loci for the $T$ samples using bioinformatics pipelines such as
e.g., BWA \citep{li2009fast}, Samtools \citep{li2009sequence}, GATK
\citep{mckenna2010genome}, etc.  Let $\bN$ and $\bn$ denote $S \times
T$ matrices of these counts,
$N_{st}$ and $n_{st}$ denoting the total number of reads and the
number of reads that bear a mutated sequence, respectively, at locus
$s$ for tissue sample $t$, $s=1, \ldots, S$ and $t=1, \ldots, T$.
Following \cite{klambauer2012cn}, we assume a Poisson sampling model
for $N_{st}$, 
\begin{eqnarray}
   N_{st} \mid \phi_t, M_{st} &\indep& \Poi(\phi_tM_{st}/2). \label{eq:like_pois}
\end{eqnarray}
Here, $M_{st}$ is the sample copy number that represents an average copy number across
subclones. We will formally define and model $M_{st}$ using subclonal copy numbers ($\bL$)
next.  $\phi_t$ is the expected number of reads in sample $t$ 
 if there were no CNV (the sample copy number equals 2).  That is, when $M_{st}=2$, the
Poisson mean becomes $\phi_t$. 

Conditional on $N_{st}$ we assume a binomial sampling model for
$n_{st}$ conditional on $N_{st}$;
\begin{eqnarray}
   n_{st} \mid N_{st}, p_{st} &\indep& \Binom(N_{st}, p_{st}). 
  \label{eq:like_binom}
\end{eqnarray}
Here $p_{st}$ is the success probability of observing a read with
a variant sequence.  It is interpreted as the
expected variant allele fractions (VAFs) in the sample.  In the following discussion we will 
represent $p_{st}$ in terms of the underlying matrices $\bL$ and $\bZ$. 

%%%%%%%%%%%%%%%%%%%%%%%%%%%%%%%%%%%%%%%%%%%%%%%%%%%%%%%%%%%%%%%%%%%%%%%%%%%%%%%%%%%
\subsection{Prior}
%%%%%%%%%%%%%%%%%%%%%%%%%%%%%%%%%%%%%%%%%%%%%%%%%%%%%%%%%%%%%%%%%%%%%%%%%%%%%%%%%%%
\paragraph*{Construction of $M_{st}$.}
Let $C$ denote the unknown number of subclones in $T$ samples. 
We first relate $M_{st}$ to CNV at locus $s$ for sample $t$.  We
construct a prior model for $M_{st}$ in two steps, using the notion
that each sample is composed of a mixture of $C$ subclones. Let
$w_{tc}$ denote the proportion of subclone $c$, $c=1, \ldots, C,$ in
sample $t$ and let $\ell_{sc} \in \{0, 1, 2, \ldots, Q\}$ denote the
number of copies at locus $s$ in subclone $c$ where $Q$ is a
pre-specified maximum number of copies.  Here $Q$ is
 an arbitrary upper bound that is used as a mathematical
device rather than having any biological meaning.   The event $\ell_{sc}=2$ 
means no copy number variant at locus $s$ in subclone $c$,
$\ell_{sc}=1$ indicates one copy loss and $\ell_{sc}=3$ indicates
one copy gain.  Then the
mean number of copies for sample $t$  can be expressed as the
weighted sum of the number of copies over $C$ latent subclones where
the weight  $w_{tc}$ denotes the  cellular fractions  of subclone
$c$ in sample $t$. We assume 
\begin{eqnarray}
     M_{st} = \ell_{s0}w_{t0} + \sum_{c=1}^C w_{tc} \ell_{sc}, 
%    \equiv \epsilon_{t0} + \sum_{c=1}^C w_{tc} z_{sc}. 
\label{eq:model_M}
\end{eqnarray}
 The second term % in \eqref{eq:model_M}, 
$\sum_{c=1}^C w_{tc} \ell_{sc}$ reflects the key assumption of
decomposing the sample copy number  into a weighted average of subclonal
copy numbers. The first term, $\ell_{s0}w_{t0}$ denotes the expected
copy number from a background subclone to account for potential noise
and artifacts in the data, labeled as subclone $c=0$. We assume no
CNVs at any the locations for the background subclone, that is,
$\ell_{s0}=2$ for all $s$.
%\cut
% In particular, $w_{t0}$ will be modeled as a small proportion of the
% background subclone in sample $t$. The prior on $\bw_t=(w_{t0},
% \ldots, w_{tC})$ is defined later.

\paragraph*{Prior on $\bL$.}
We develop a feature-allocation  prior for a latent random matrix of
copy numbers, 
$\bL=[\ell_{sc}]$, % where $\ell_{sc} \in \{0, 1, 2, \ldots, Q\}$ for
$c=1, \ldots, C$ and $s=1, \ldots, S$. 
 We first construct a prior $p(\bL \mid C)$ conditional on $C$. 
Let $\bpi_c = (\pi_{c0},
\pi_{c1}, \ldots, \pi_{cQ})$ where $p(\ell_{sc}=q) = \pi_{cq}$ and
$\sum_{q=0}^Q \pi_{cq}=1$.  As a prior distribution of $\bpi_c$, we
use a beta-Dirichlet distribution developed in \cite{kim2012bayesian}.
Conditional on $C$, $p(\ell_{sc} \neq 2)=(1-\pi_{c2})$
follows a beta distribution with parameters, $\alpha/C$ and $\beta$
and $\bpit=(\pit_{c0}, \pit_{c1}, \pit_{c3}, \ldots, \pit_{cQ})$,
where $\pit_{cq} = \pi_{cq}/(1-\pi_{c2})$ with $q\ne 2$, follows a
Dirichlet distribution with parameters, $(\gamma_0, \gamma_1,
\gamma_3, \ldots, \gamma_Q)$.  Assuming a priori independence among
subclones, we write $\bpi_c \iid \BeDir \allowbreak (\alpha/C, \beta,
\gamma_0, \gamma_1, \gamma_3, \ldots, \gamma_Q)$.    For $\beta=1$, the marginal limiting distribution of $\bL$ can be shown
to define a categorical Indian buffet process (cIBP) as $C
\rightarrow \infty$ 
\citep{Subhajit:2013,Sengupta:2014}.    

\paragraph*{Construction of $p_{st}$ and prior on $\bZ$.}  
To model the expected VAF of the sample, $p_{st}$, we construct
another feature allocation model linking $p_{st}$ with $\ell_{sc}$. 
We introduce an $S \times C$ matrix, $\bZ$ whose entries,
$z_{sc} \in \{0, \ldots, \ell_{sc}\}$ denote the number
$z_{sc} \leq \ell_{sc}$ of alleles bearing a variant sequence among
the total of $\ell_{sc}$ copies at locus $s$ in 
 subclone $c$. Assume $z_{sc}=0$ if $\ell_{sc}=0$, and given $\ell_{sc} > 0$,
\begin{eqnarray}
  z_{sc} \mid \ell_{sc} \sim \DU(0, 1, \ldots, \ell_{sc}), \label{eq:prior_z}
\end{eqnarray}
where $\DU(\cdot)$ indicates a discrete uniform distribution. 

Next, we write $p_{st}$ 
in \eqref{eq:like_binom} 
as a ratio between the expected number of variant alleles and the
expected sample copy number. In particular, the expected number of
variant alleles is a weighted sum of subclonal variant allele counts over $(C+1)$ latent subclones
including the background subclone, and the expected VAF is
\vskip -.4in
{\small
\begin{eqnarray}
    p_{st} = \frac{p_0 z_{s0} w_{t0} + \sum_{c=1}^C w_{tc} z_{sc}}{M_{st}}  \label{eq:model_p}
\end{eqnarray}
}
\vskip -.25in
\noindent
Similar to the previous argument for \eqref{eq:model_M}, the term
$\sum_{c=1}^C w_{tc} z_{sc}$ in \eqref{eq:model_p} reflects the assumption that the
sample-level variant allele count is a weighted average of subclonal
variant allele counts. The first term of the numerator, $p_0 z_{s0}
w_{t0}$  describes the background subclone and experimental
noise. Specifically, we let $z_{s0}=2$ for all $s$ to denote the
number of variant alleles in the background subclone. %\cut
% and $w_{t0}$ in
% \eqref{eq:model_p} accounts   for   the small cellular fraction of the
% background subclone.  
We add a global parameter $p_0$ to account for
artifacts and experimental noise that would produce variant reads
even if no subclones were to possess variant alleles. Since $p_0$ does
not depend on $s$ or $t$, it can be estimated by pooling data from all
loci and samples, and does not affect the identifiability of the model.
We consider $p_0 \sim \Be(a_{00}, b_{00})$ with $a_{00} \ll b_{00}$ to
inform a small $p_0$ value {\it a priori}.  Equation
\eqref{eq:model_p} echos our previous discussion for Figure
\ref{fig:data_1}(b), modeling the VAFs as a mixture of subclonal
variant alleles.

% For each locus $s$ and subclone $c$
% let $z_{sc} \in \{0, \ldots, \ell_{sc}\}$ denote the
% number of copies among the $\ell_{sc}$ copies that are
% bearing a variant sequence.
%  Then the expected copy number with variant allele among the $M_{st}$
% copies can be expressed as
% $m_{st} = p_0z_{s0}w_{t0} + \sum_{c=1}^C w_{tc} z_{sc}$. As before we
% added an additional term for a background subclone $c=0$. 
% This implies that the expected VAF is 
% \begin{eqnarray}
%     p_{st} = \frac{m_{st}}{M_{st}}=\frac{p_0  z_{s0}   w_{t0} + \sum_{c=1}^C w_{tc} z_{sc}}{M_{st}}. 
% \label{eq:model_p}
% \end{eqnarray}
% We fix $z_{s0}=2$ for all $s$ and add a global parameter $p_0$ to
% account for artifacts and experimental noise that would produce
% variant reads even if the subclones are not variant. 
% Since $p_0$ does not depend on $s$ or
% $t$, it can be estimated pooling data from all loci and samples and
% does not affect the identifiability of the model.   We assume $p_0
% \sim \Be(a_{00}, b_{00})$ with $a_{00} \ll b_{00}$ to 
%  shrink $p_0$ to small values a priori. 
% In words, $p_{st}$, the proportion of
% reads with SNV at locus $s$ in sample $t$ is the expected proportion
% of alleles having SNV.   
% We continue the model construction with a prior on $z_{sc}$.
% Given $\ell_{sc}$, we assume that $z_{sc}$ is equally likely to be
% any integer in $\{0, \ldots, \ell_{sc}\}$. 
% That is, 
% $$
%    p(z_{sc}= z \mid \ell_{sc}) = \frac{1}{\ell_{sc}+1}
% $$
% and $z_{sc}=0$ if $l_{sc}=0.$ 

\paragraph*{Prior for $\bw$.}
Next, we introduce a prior distribution for the weights $w_{tc}$ 
in \eqref{eq:model_M} and \eqref{eq:model_p}.
The subclones are common for all tumor samples, but the relative
weights $w_{tc}$ vary across   tumor   samples.  We assume
independent Dirichlet priors as follows.
Let $\theta_{tc}$
denote an (unscaled) abundance level of subclone $c$ in tissue sample
$t$.  We assume $\theta_{tc} \mid C \iid \Ga(d, 1)$ for $c=1, \ldots,
C$ and  
$\theta_{t0} \iid \Ga(d_0, 1)$.  We then define
$$
   w_{tc}= \theta_{tc}/\sum_{c^\prime=0}^{C} \theta_{tc^\prime},
$$
as the relative weight of subclone $c$ in sample $t$.  
This is equivalent to $\bw_t \mid C \iid \Dir(d_0, d, \ldots, d)$ for $t=1, \ldots,
T$.  Using $d_0 < d$ implies that the background subclone takes a
smaller proportion in a sample. 

Finally, we complete the model construction with a prior on
the unknown number of latent subclones $C$. 
%The number of latent subclones, $C$, is unknown.  
We use a geometric distribution, $C \sim \Geom(r)$ where
$\mathrm{E}(C)=1/r$.  Conditional on $C$, the two latent matrices,
$\bL$ and $\bZ$ describes $C$ latent tumor subclones
that are thought of composing the observed samples and $\bw_t$ provides the relative proportions over those $C$ subclones in sample $t$.
Joint inference on $C$, $\bL$, $\bZ$ and $\bw_t$
explains tumor heterogeneity.  

%\note{moved this par here from above -- P}
The construction of the subclones, including the number of subclones,
$C$, the subclonal copy number, $\ell_{sc}$, and the number of copies
having SNV, $z_{sc}$ are latent. The subclones are not directly
observed. They are only defined as the components of the assumed
mixture that gives rise to the observed CNV and VAFs.  The key terms,
$\sum_{c=1}^C w_{tc}\ell_{sc}$ in \eqref{eq:model_M} and $\sum_{c=1}^C
w_{tc}z_{tc}/M_{st}$ in \eqref{eq:model_p} allow us to indirectly
infer subclones by explaining $M_{st}$ and $p_{st}$ as arising from
sample $t$ being composed of a mix of hypothetical subclones which
have $\ell_{sc}$ copies of which $z_{sc}$ actually carry a variant 
at locus $s$.

Lastly, we take account of different average read counts in $T$ samples through $\phi_t$. $\phi_t$ represents the expected read count with two copies in sample $t$ and assume $\phi_t \indep \Ga(a_t, b_t)$ where $\mathrm{E}(\phi_t)=a_t/b_t$. % Recall \eqref{eq:like_pois}, $N_{st}$ is determined by the average read count in sample $t$, $\phi_t$ and latent copy number at position $s$ in sample $t$, $M_{st}$, in words.
% \note{Consider modeling $\phi_t$ using informative priors. They can be easily elicited using the entire NGS data. --Y  I agree. We have fairly different results for the lung cancer data (Juhee)}

%%%%%%%%%%%%%%%%%%%%%%%%%%%%%%%%%%%%%%%%%%%%%%%%%%%%%%%%%%%%%%%%%%%%%%%%%%
\subsection{Posterior Simulation}
\label{sec:post_sim}
%%%%%%%%%%%%%%%%%%%%%%%%%%%%%%%%%%%%%%%%%%%%%%%%%%%%%%%%%%%%%%%%%%%%%%%%%%
Let  $\bx = (\bL, \bZ, \bth, \bphi, \bpi, p_0)$ denote all unknown parameters, where $\bth = \{\theta_{tc}\}$ and $\bpi = \{\pi_{cq}\}$.  We implement inference via posterior Markov chain Monte Carlo (MCMC) simulation. That is, by generating a Monte Carlo sample of
$\bx_i \sim p(\bx \mid \bn, \bN)$, $i=1,\ldots,I$. 
MCMC posterior simulation  proceeds by sequentially
using transition probabilities that update a subset of parameters at a
time. See, for example \cite{brooks2011handbook} for a review. 
 
For fixed $C$ such MCMC simulation is straightforward. 
Gibbs sampling transition probabilities
are used to update $\ell_{sc}$, $z_{sc}$, $\pi_{cq}$ and $\phi_t$ and
Metropolis-Hastings transition probabilities are used to update
$\btheta$ and $p_0$.  It is possible to improve the mixing of the
Markov chain by updating all columns in row $s$ of the matrices $\bL$
and $\bZ$ jointly by means of a Metropolis-Hastings transition
probability that proposes changes in the entire row vector 
$\bz_s$ and $\bell_s$.

The construction of transition probabilities that involves a change of
$C$ is more difficult, since the dimension of $\bL$, $\bZ$, $\bpi$
and $\bth$ changes as $C$ varies. We use the approach proposed in
\cite{Lee2014TH}  for posterior simulation in a similar model. 
We split the data into a small training set $(\bn',\bN')$  
with $n_{st}^\prime=b_{st} n_{st}$, $N'_{st}=b_{st} N_{st}$,
and a test data set, $(\bn^{''},\bN^{''})$ with $n^{''}_{st}=(1-b_{st})
n_{st}$ etc. In the implementation we use $b_{st}$ generated from
$\Be(25, 975)$ for the simulation studies and $\Be(30, 970)$ for the lung cancer data.  
 We found that using a random $b_{st}$ worked better than a fixed
fraction $b$ across all samples and loci. 
Let $p_1(\bx \mid C) = p(\bx \mid \bN^\prime,
\bn^{\prime},C)$ denote the posterior distribution under $C$ using the
training sample. We use $p_1$ in two instances. First, we replace the
original prior $p(\bx \mid C)$ by $p_1(\bx \mid C)$ and, second, we
use $p_1(\cdot)$ as proposal distribution $q(\bxt \mid \Ct) = p_1(\bxt \mid
\Ct)$ in a reversible jump (RJ) style transition probability where $\Ct$ is a proposed value of $C$.
The test data is then used to evaluate the acceptance
probability.
The critical advantage of using the same $p_1(\cdot)$ as prior
and proposal distribution is that the normalization constant cancels
out in the Metropolis-Hastings acceptance probability. 

We summarize the joint posterior distribution, $p(C, \bL, \bZ, \bpi,
\bphi, \bw, p_0 \mid \bn, \bN)$ by factorizing it as  
\begin{multline}
p(C \mid \bn, \bN)\,
p(\bL \mid \bn, \bN, C)\,
p(\bZ, \bpi \mid \bn, \bN, C, \bL)\,
p(\bw \mid \bL, \bZ, \bn, C)\, 
p(\bphi, p_0 \mid \bn,\bN, C).  
\nonumber
\end{multline}
Using the posterior Monte Carlo sample we (approximately)
evaluate the marginal posterior $p(C \mid \bn,\bN)$ and determine the
maximum a posteriori (MAP) estimate $\Cs$.  We follow
\cite{Lee2014TH} to define $\bL^\star$ conditional on $\Cs$.  For
any two $S \times \Cs$ matrices, $\bL$ and $\bL^\prime$, $1 \le c,
c^\prime \le \Cs$, let $\DD_{cc'}(\bL,\bL^\prime)=\sum_{s=1}^S
|\ell_{sc} - \ell^\prime_{sc^\prime}|$. 
We then define a distance 
$
   d(\bL, \bL^\prime) = 
      \min_{\bsig} \sum_{c=1}^{\Cs} \DD_{c,\sig_c}(\bL,\bL^\prime),
$
where $\bsig = (\sig_1,\ldots,\sig_C)$ is a permutation of
$\{1, \ldots, 
\Cs\}$ and the minimum is over all possible permutations. A posterior
point estimate for $\bL$ is defined as
$$
  \bL^\star = \arg \min_{\bL^\prime} 
  \int d(\bL, \bL^\prime)\,  dp(\bL \mid \bn, \bN, \Cs) \approx 
  \arg \min_{\bL^\prime}  \sum_{i=1}^I d(\bL^{(i)}, \bL^\prime),  
$$
for a posterior Monte Carlo sample, $\{\bL^{(i)}, i=1, \ldots, I\}$.   We report posterior point estimates $\bZ^\star$, $\bw^\star$ and $\bpi^\star$ conditional on $\Cs$ and $\bL^\star$.    Finally, we report $\bphi^\star$ and $p_0^\star$ as the posterior mean of $\bphi$ and $p_0$ conditional on $\Cs$.

%%%%%%%%%%%%%%%%%%%%%%%%%%%%%%%%%%%%%%%%%%%%%%%%%%%%%%%%%%%%%%%%%%%%%%%%%%%%%%%%%%%
\section{Simulation}
\label{sec:Simulation}
%%%%%%%%%%%%%%%%%%%%%%%%%%%%%%%%%%%%%%%%%%%%%%%%%%%%%%%%%%%%%%%%%%%%%%%%%%%%%%%%%%%
%%%%%%%%%%%%%%%%%%%%%%%%%%%%%%%%%%%%%%%%%%%%%%%%%%%%%%%%%%%%%%%%%%%%%%%%%%%%%%%%%%%
\subsection{Simulation 1}
\label{sec:Simulation0}
\begin{figure}[ht!]
  \begin{center}
\begin{tabular}{ccc}
   \includegraphics[width=.32\textwidth]{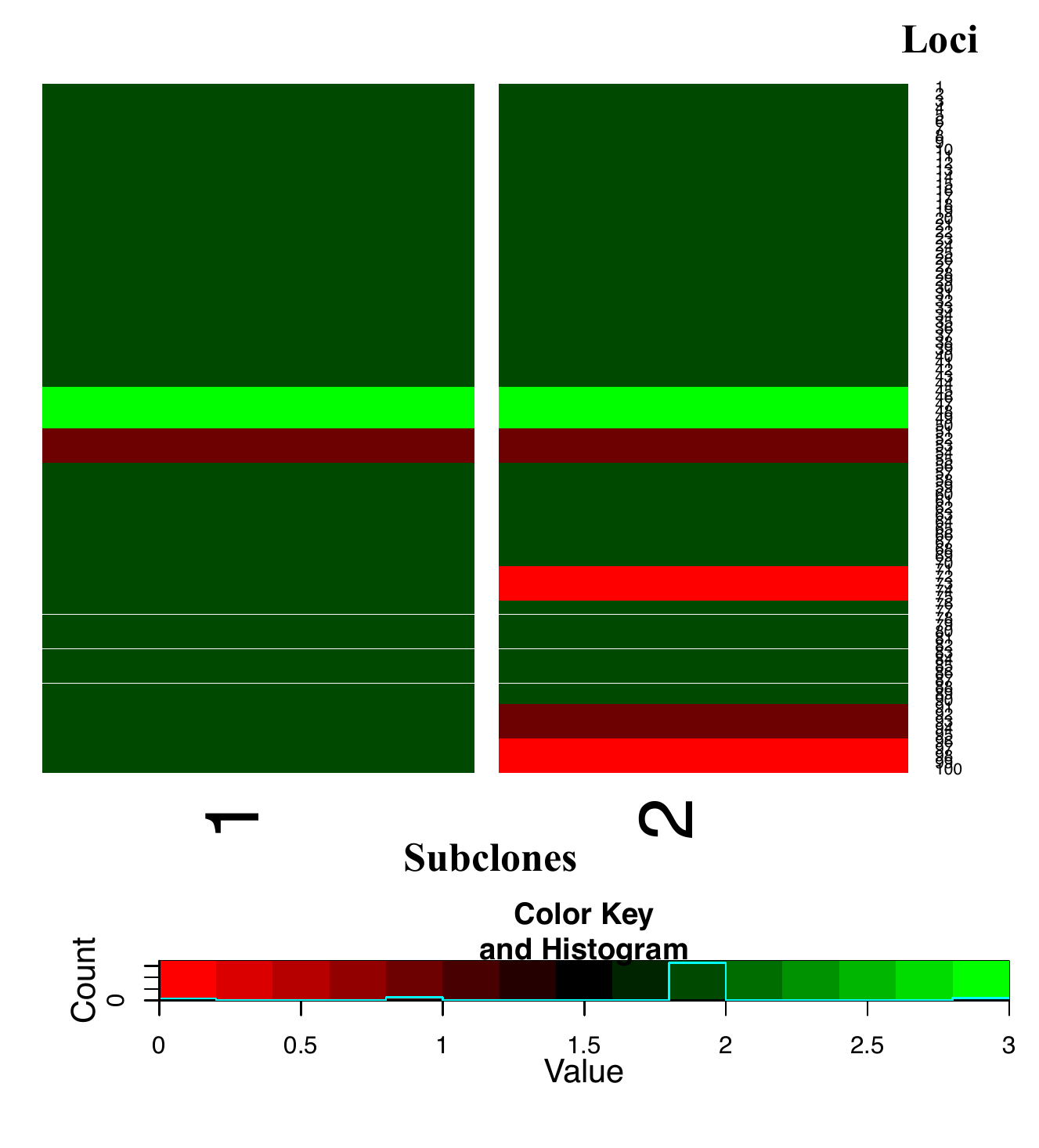} &
   \includegraphics[width=.32\textwidth]{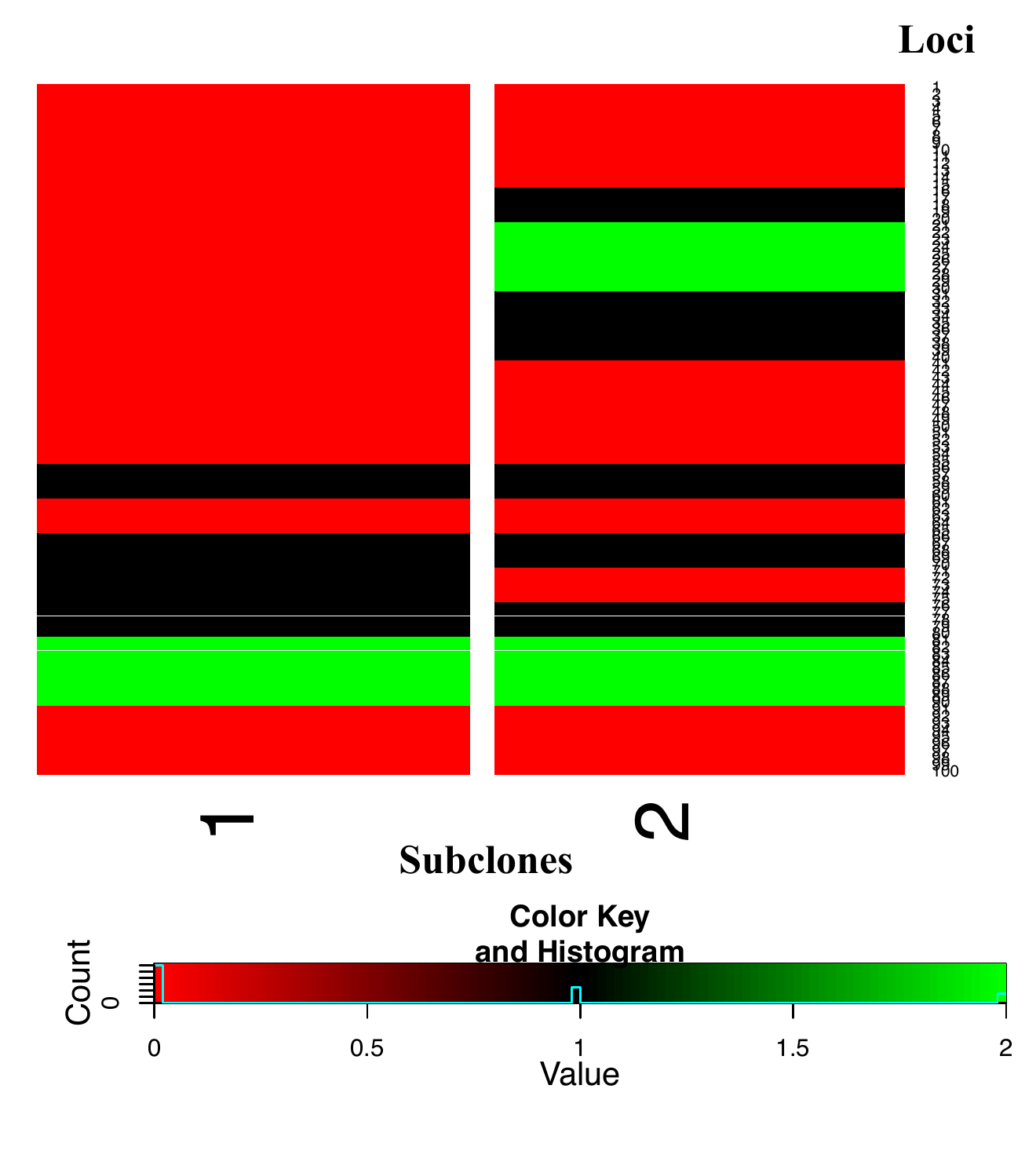} &
   \includegraphics[width=.32\textwidth]{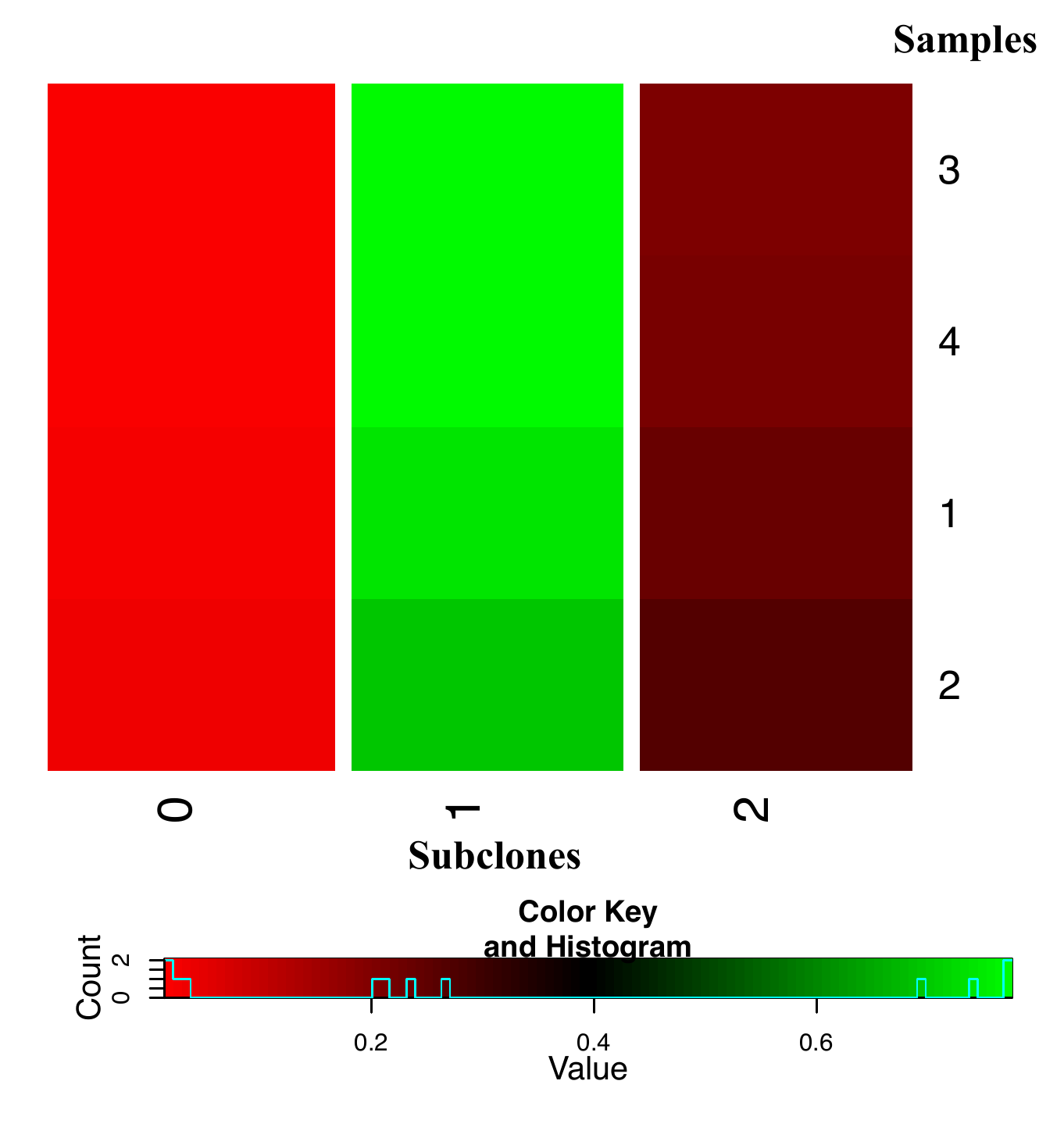} \\
  (a) $\bL^\true$ &
  (b) $\bZ^\true$ &
  (c) $\bw^\true$ \\ 
\end{tabular}
 \end{center}
\caption{Simulation 1: simulation truth.}
\label{fig:Sim0_tr}
\end{figure}

%%%%%%%%%%%%%%%%%%%%%%%%%%%%%%%%%%%%%%%%%%%%%%%%%%%%%%%%%%%%%%%%%%%%%%%%%%%%%%%%%%%
We assess the proposed model via simulation.  We generate
 hypothetical
read counts
for a set of $S=100$ loci in $T=4$ hypothetical samples.
In the simulation truth, we assume two latent subclones
($C^{\true}=2$) as well as a background subclone ($c=0$) with all SNVs
bearing variant sequences with two copies. We use $Q=3$.  The
simulation truth $\bL^\true$ is shown in Figure~\ref{fig:Sim0_tr}(a)
where green color (light grey) in the panels indicates a copy gain
($\ell_{sc}=3$) and red color (dark grey) indicates two copy loss
($\ell_{sc}=0$).  Panel (b) shows the simulation truth $\bZ^\true$.
Similar to $\bL^\true$, green color indicates three copies with SNV
and red color indicates zero copies with SNV. We generate
$\phi^\true_t \iid \Ga(600, 3)$, $t=1, \ldots, 4$ and then
generate $\bw^\true \iid \Dir(0.4, 30.0, 10.0)$.  The weights $\bw^\true$
are shown in Figure~\ref{fig:Sim0_tr}(c). Similar to the other
heatmaps, green color (light grey) in panel (c) represents high
abundance of a subclone in a sample and red color (dark grey) shows
low abundance. On average, subclone 1 takes $w_{tc}$ close to 0.75 for
all the samples, with little heterogeneity across samples. 
Using the assumed
$\bL^\true$, $\bZ^\true$ and $\bw^\true$ and letting $p_0^\true=0.05$,
we generate $N_{st} \sim \Poi(\phi_t^\true M_{st}^\true/2)$ and
$n_{st} \sim \Binom(N_{st}, p^\true_{st})$.

%%%%%%%%%%%%%%%%%%%%%%%%
%%%%%%%%%%%%%%%%%%%%%%%%%

To fit the proposed model, we fix the hyperparameters
as $r=0.2$, $\alpha=2$, $\gamma_q=0.5$
for $q=0, 1, 3(=Q)$,  $d_0=0.5$, $d=1$, $a_{00}=0.3$ and
$b_{00}=5$. For the prior on $\phi_t$, we let $b=3$ and specify $a$ by setting
the median of the observed $N_{st}$ to be the prior mean.  For each value of $C$, we
initialized $\bZ$ using the observed sample proportions and $\bL$
using the initial $\bZ$.  We generated initial values for
$\btheta_{tc}$ and $p_0$ by prior draws. We generated $b_{st} \iid
\Be(25, 975)$ to construct the training set and ran the MCMC
simulation over 16,000 iterations, discarding the first 6,000
iterations as initial burn-in.

%%%%%%%%%%%%%%%%%%%%%%%%
\begin{figure}[ht!]
  \begin{center}
\begin{tabular}{ccc}
   \includegraphics[width=.32\textwidth]{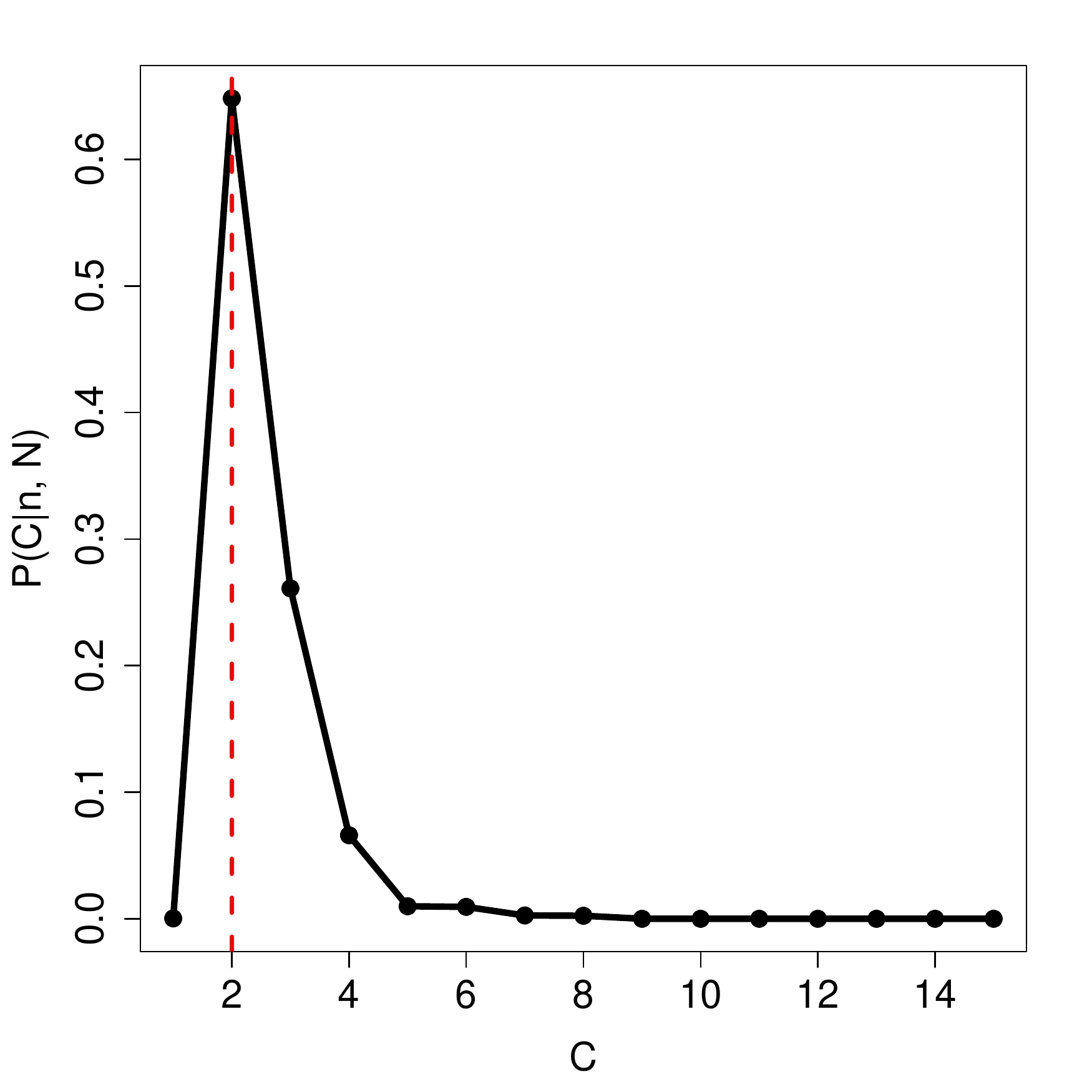} &
   \includegraphics[width=.32\textwidth]{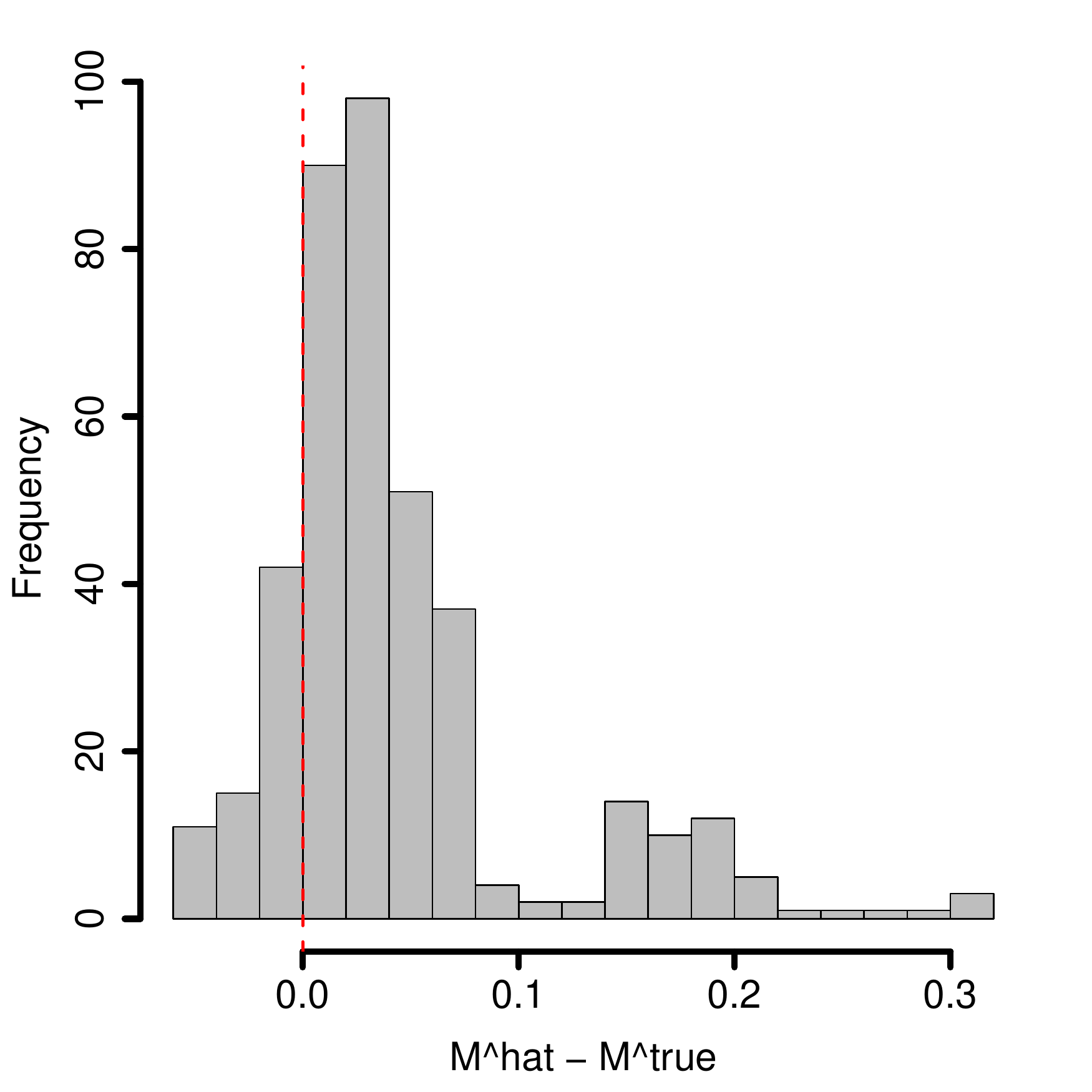} &
   \includegraphics[width=.32\textwidth]{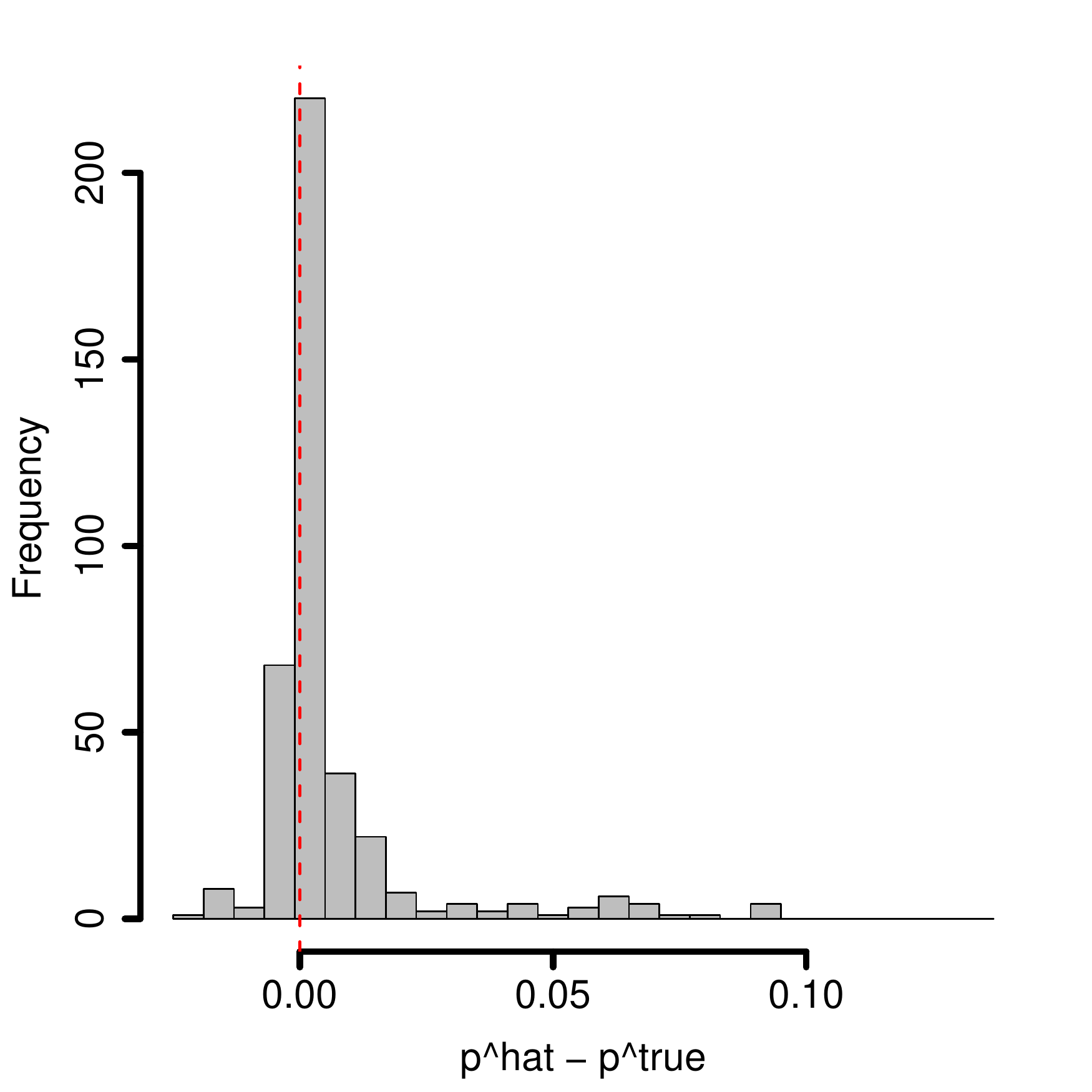} \\
  (a) $p(C \mid \bn,\bN)$ &
  (b) $\hat{M}_{st}- M^\true_{st}$ &
  (c) $\hat{p}_{st}- p^\true_{st}$ \\
   \includegraphics[width=.32\textwidth]{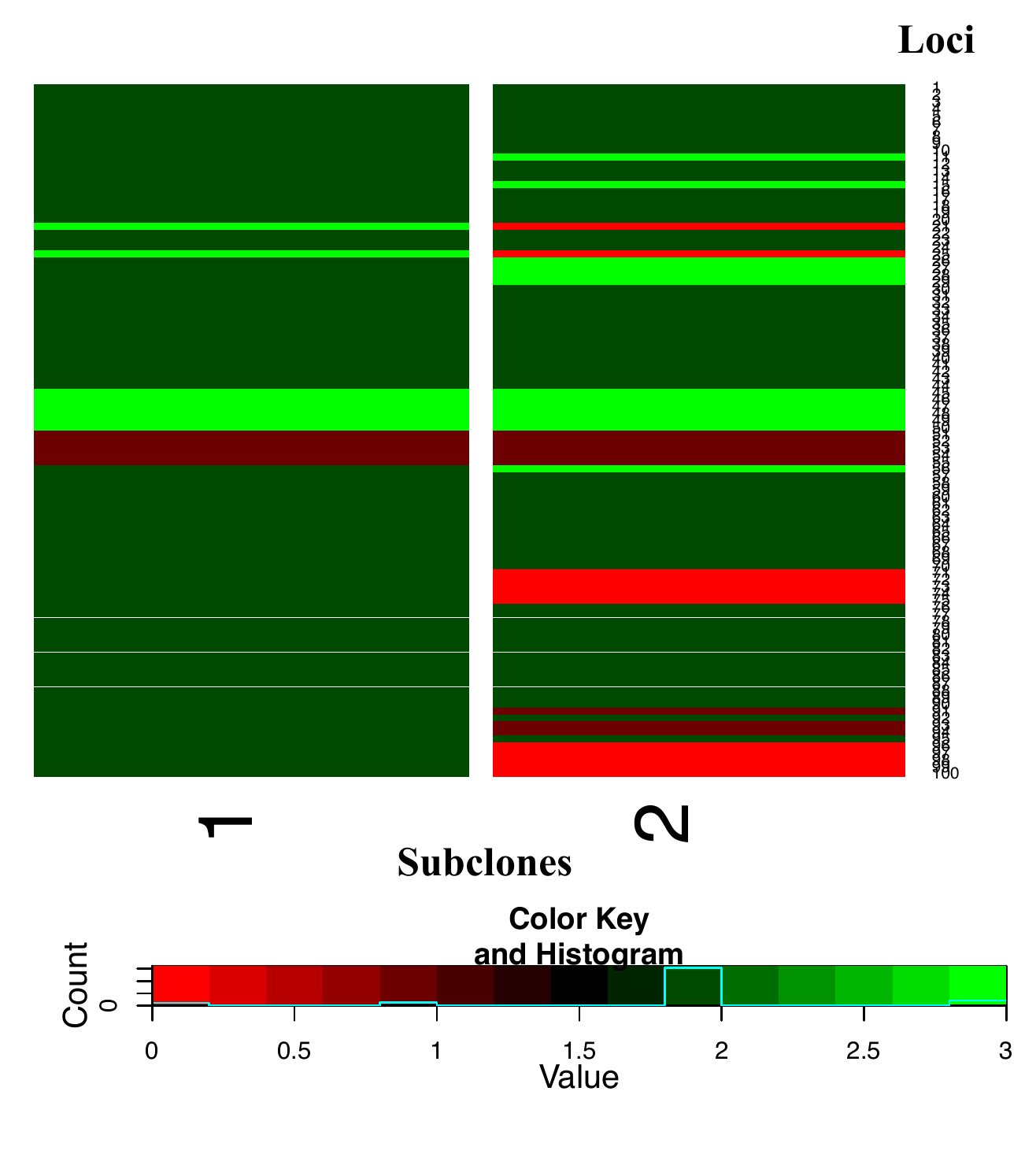} &
   \includegraphics[width=.32\textwidth]{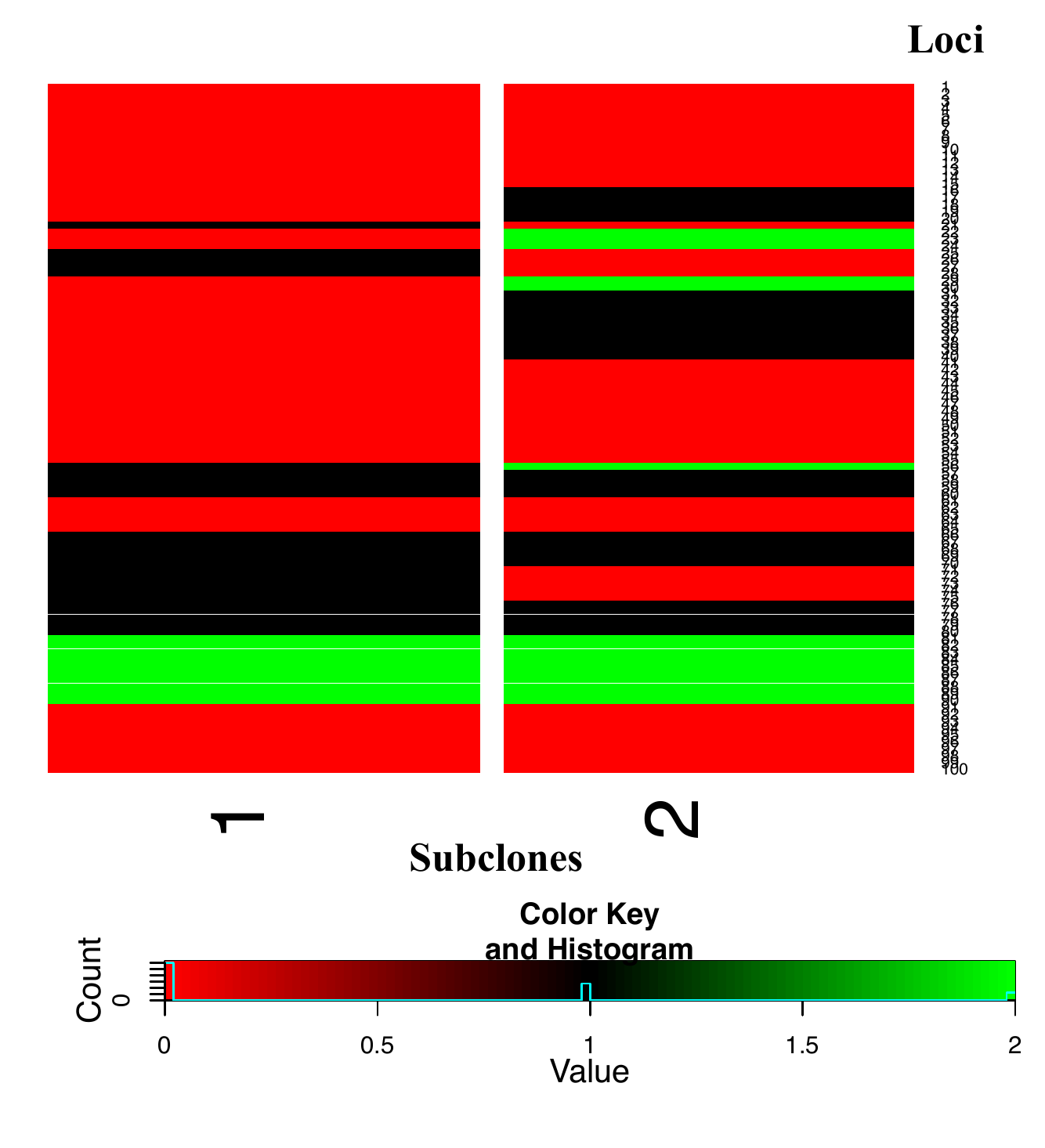} &
   \includegraphics[width=.32\textwidth]{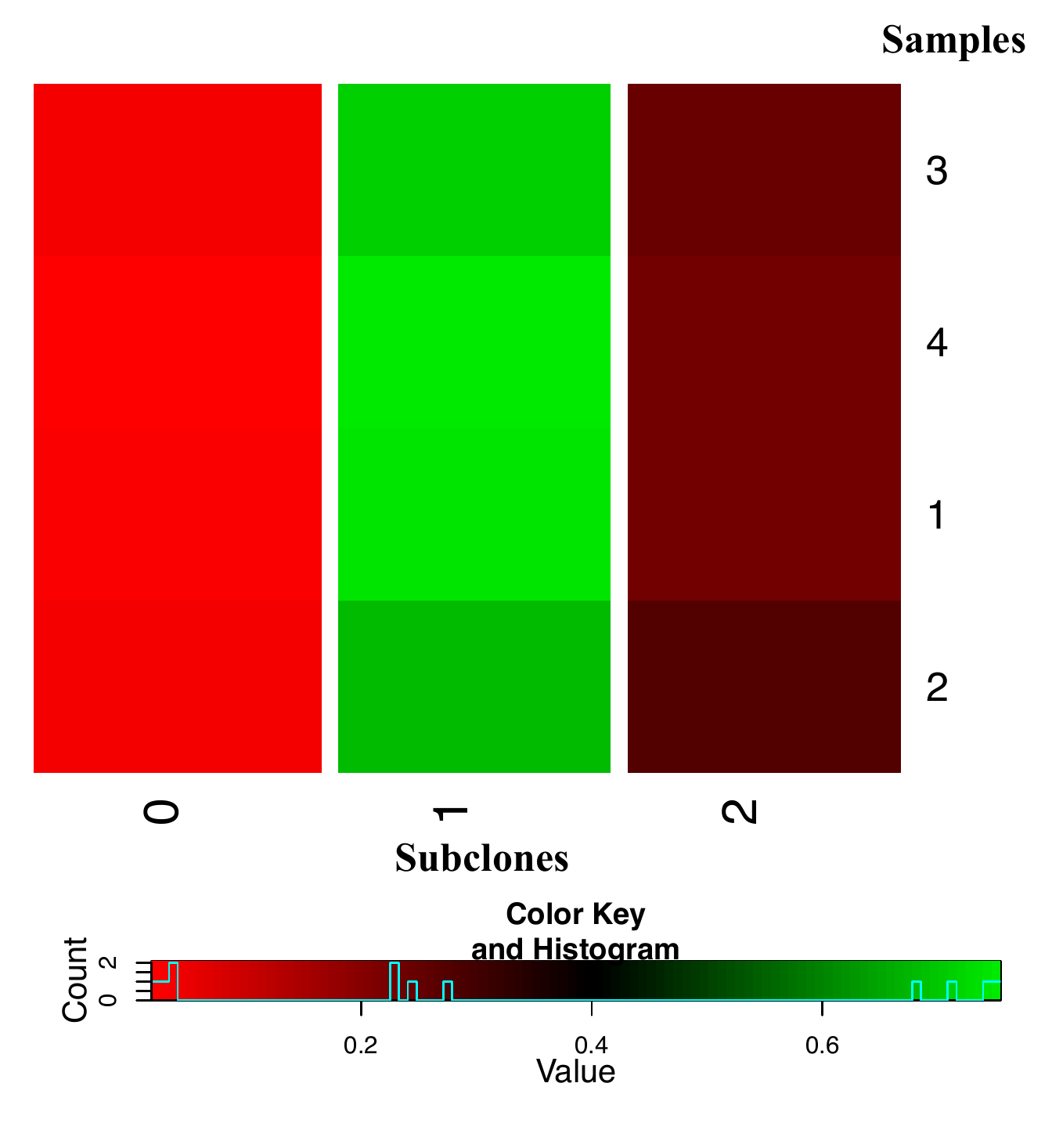} \\
  (d) $\bL^\star$ &
  (e) $\bZ^\star$ &
  (f) $\bw^\star$ \\
\end{tabular}
 \end{center}
\caption{Posterior inference for Simulation 1.}
\label{fig:Sim0_post}
\end{figure}
%%%%%%%%%%%%%%%%%%%%%%%%%

% ***************************************************************
% inference for SIM 1 
% ***************************************************************
Figure~\ref{fig:Sim0_post}(a) shows 
$p(C \mid \bn, \bN)$. 
The dashed vertical line marks the simulation truth
$C^{\true}=2$.  The posterior mode $\Cstar=2$ recovers the truth.
Panels (d) through (f) show the posterior point estimates,
$\bL^\star$, $\bZ^\star$ and $\bw^\star$.  Compared to the simulation
truth in Figure~\ref{fig:Sim0_tr}, the posterior estimate recovers
subclone 1 with high accuracy, but  $\bell^\star_c$ for
subclone $c=2$ shows some discrepancies with the simulation truth. 
This is due to
small $w^\true_{tc}$, $c=2$, across all four samples
(last column in Figure~\ref{fig:Sim0_tr}c).  The discrepancy between
$\bell_2^\star$ and $\bell_2^\true$ is related to
the misspecification of $\bz^\star_{c}$ under $c=2$. 
Conditional on $\Cs$, we computed $\hat{M}_{st}$ and $\hat{p}_{st}$ and compared
to the true values. Figure~\ref{fig:Sim0_post}(b) and (c) show 
a good fit under the model for a majority of loci and samples
although the histograms include 
a small pocket of differences between the true values and their
estimates on the right tail, also possibly due to the 
misspecification of $\bell_2$ and $\bz_2$.  This simulation study illustrates that the proposed model reasonably recovers the simulation truth even with a small number of samples when the underlying structure is not complex. 

%%%%%%%%%%%%%%%%%%%%%%%%
\begin{figure}[ht!]
  \begin{center}
\begin{tabular}{cc}
   \includegraphics[width=.32\textwidth]{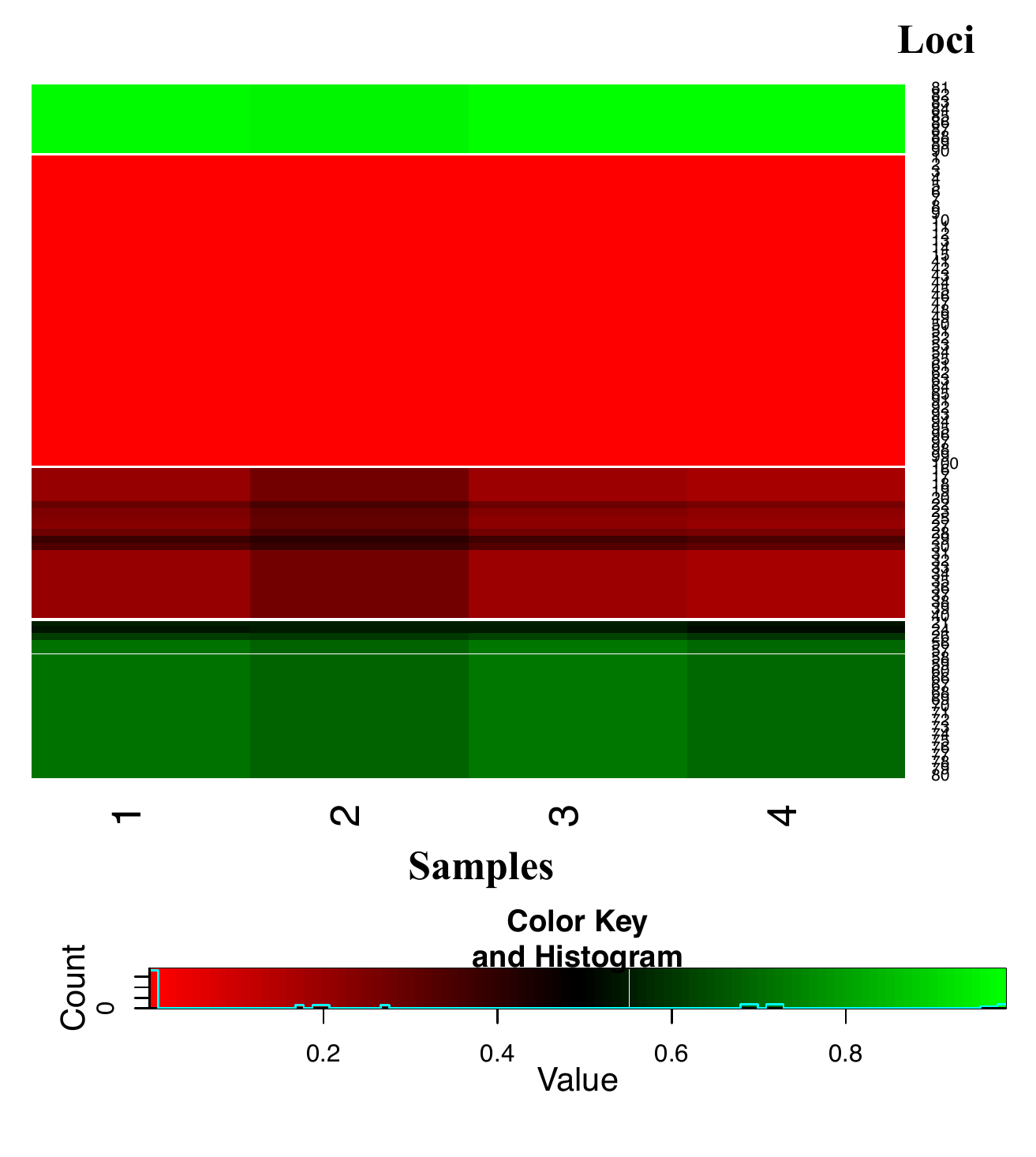} &
   \includegraphics[width=.32\textwidth]{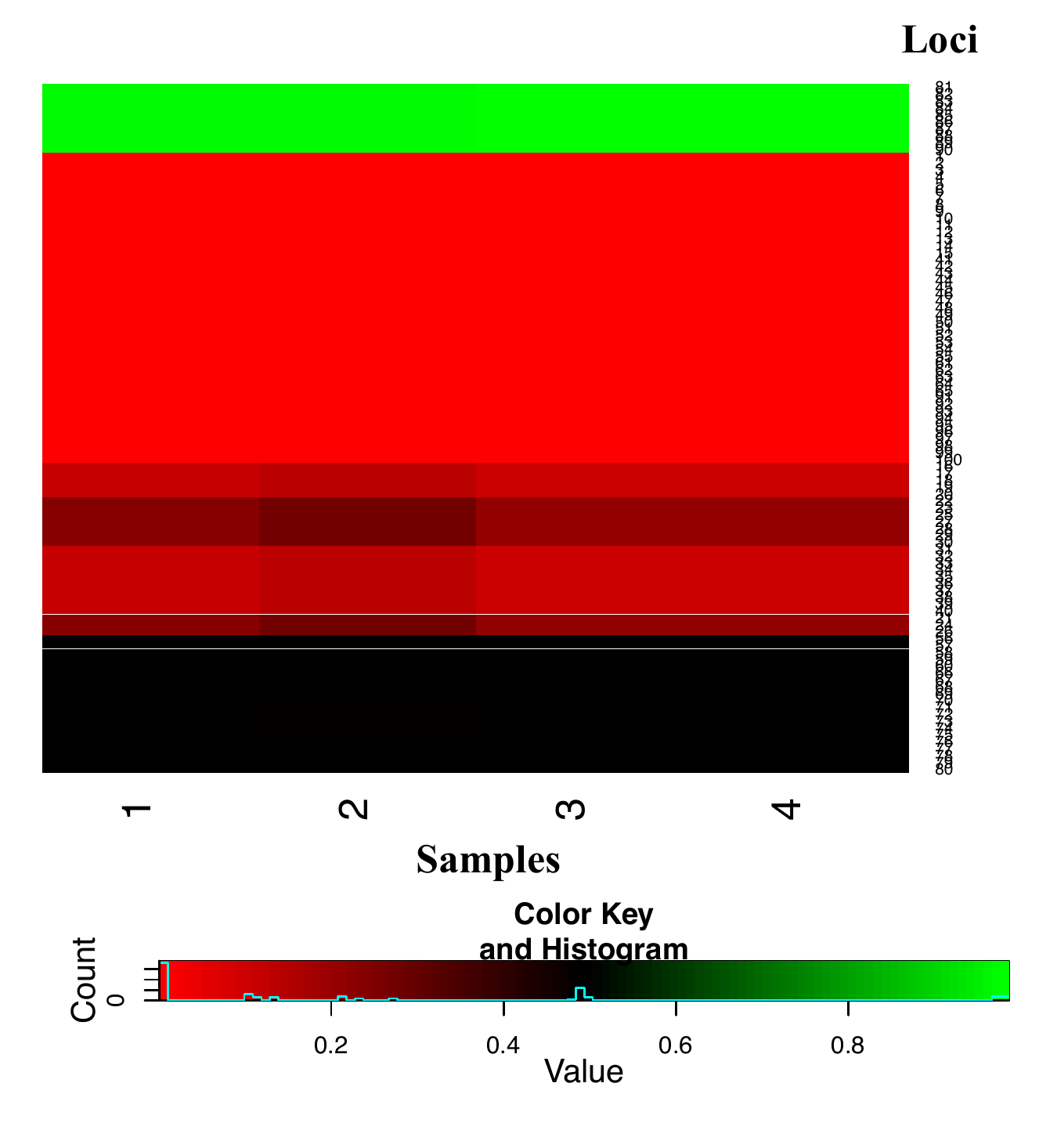} \\
  (a) Cellular prevalences &
  (b) $p^\true_{st}$ \\
\end{tabular}
 \end{center}
\caption{Heatmaps of estimated cellular prevalences from PyClone (a) and $p_{st}^\true$ (b) for Simulation 1.}
\label{fig:Sim0_pyclone}
\end{figure}
%%%%%%%%%%%%%%%%%%%%%%%%%

For comparison, we implemented PyClone \citep{roth2014pyclone} with
the same simulated data. We let the normal copy number, the minor
parental copy number and the major parental copy number be 2, 0 and
3, respectively, at each locus.  PyClone considers copy number changes
and estimates the variant allelic prevalence (fraction of clonal
population having a mutation) at a locus in a sample. 
The interpretation of variant allelic prevalences,
referred to as ``cellular prevalences'' in PyClone, 
is similar to that of $p_{st}$ in the proposed model.
%\note{{\bf Juhee:} you had "forumulation'' -- do you mean the actual  construction, as in eq (4), or the interpretation? I thought the latter..}
PyClone uses a
Dirichlet process model to identify a (non-overlapping) clustering of
the loci based on their cellular prevalences. Cellular prevalences
over loci and samples may vary but the clustering of loci is shared by
samples. Figure~\ref{fig:Sim0_pyclone}(a) shows posterior
estimates of the cellular prevalences (by color and grey shade)  and mutational
clustering (by separations with white horizontal lines) under
PyClone. Panel (b) of the figure shows a heatmap of $p_{st}^\true$.
The loci (rows) of the two heatmaps are re-arranged in the same order
for easy comparison. By comparing the two heatmaps, the cellular
prevalence estimates under PyClone are close to $p_{st}^\true$ and
 lead to a reasonable estimate of  a clustering of the loci.  However,
PyClone does not attempt to construct a description of subclones with
genomic variants.     

%\note{Juhee, as it stands, it is not obvious to me what the
%  differences are between Sim 2 and Sim 1. Why do we want to present
%  Sim 2? Also, do we have results when sample size drops to, say 4 or
%  5? That would be a good alternative example even if the results are
%  not too good. Let's talk. --Y}
%\note{{\bf J \& Y:} agreed with Y. Sim 2 might just be too much. It's
%  difficult for a reader to read through such a lengthy description of
%  sim results. I suggest we skip it (and include it later only if refs
%  bitch? I removed it (still there as ``CNV10-sim2.tex'' and renamed
%  ``Sim 3'' into ``Sim 2'' -- P} 

%%%%%%%%%%%%%%%%%%%%%%%%%%%%%%%%%%%%%%%%%%%%%%%%%%%%%%%%%%%%%%%%%%%%%%%%%%%%%%%%%%%
\subsection{Simulation 2}
\label{sec:Simulation2}
\begin{figure}[ht!]
  \begin{center}
\begin{tabular}{ccc}
   \includegraphics[width=.32\textwidth]{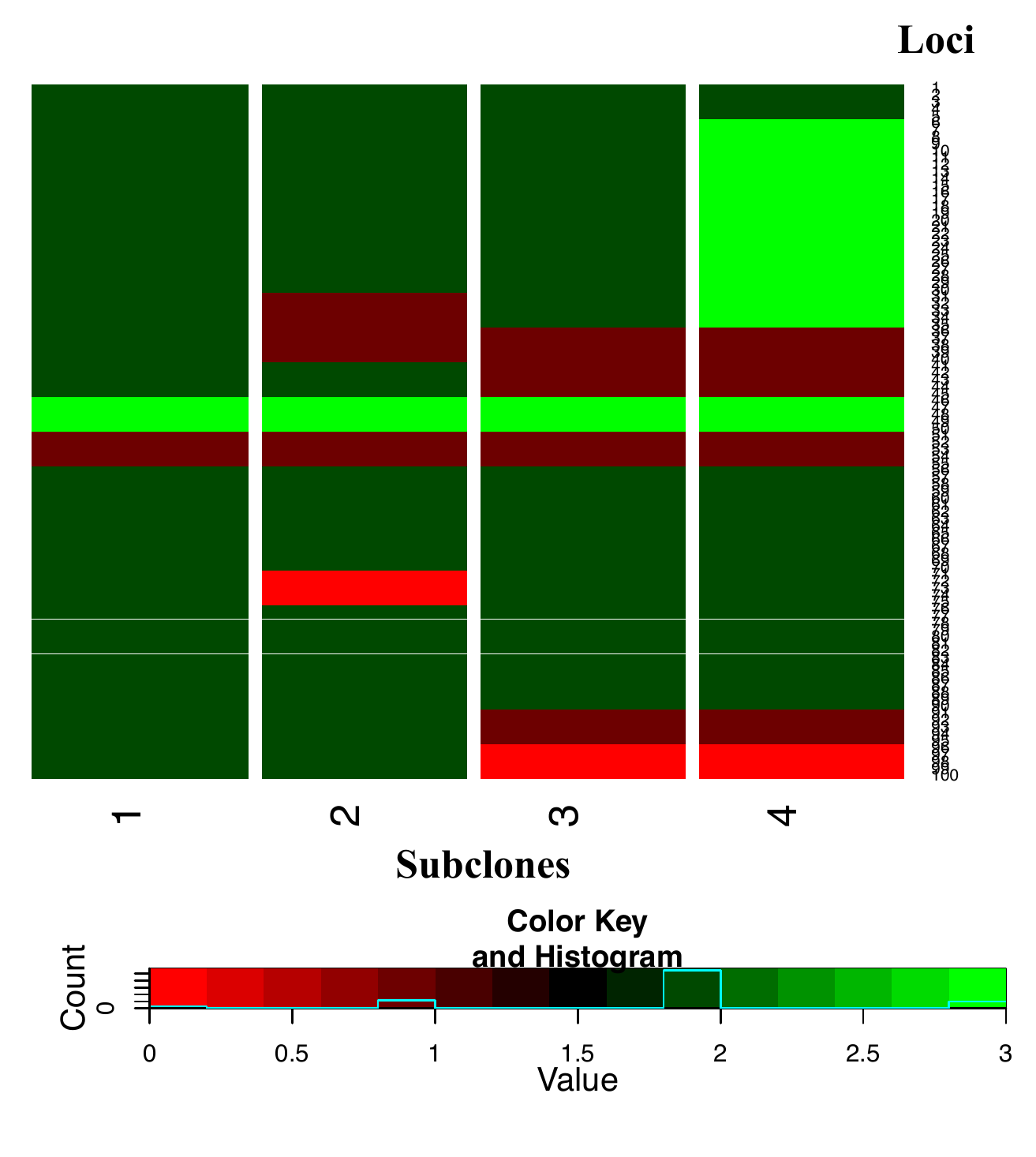} &
   \includegraphics[width=.32\textwidth]{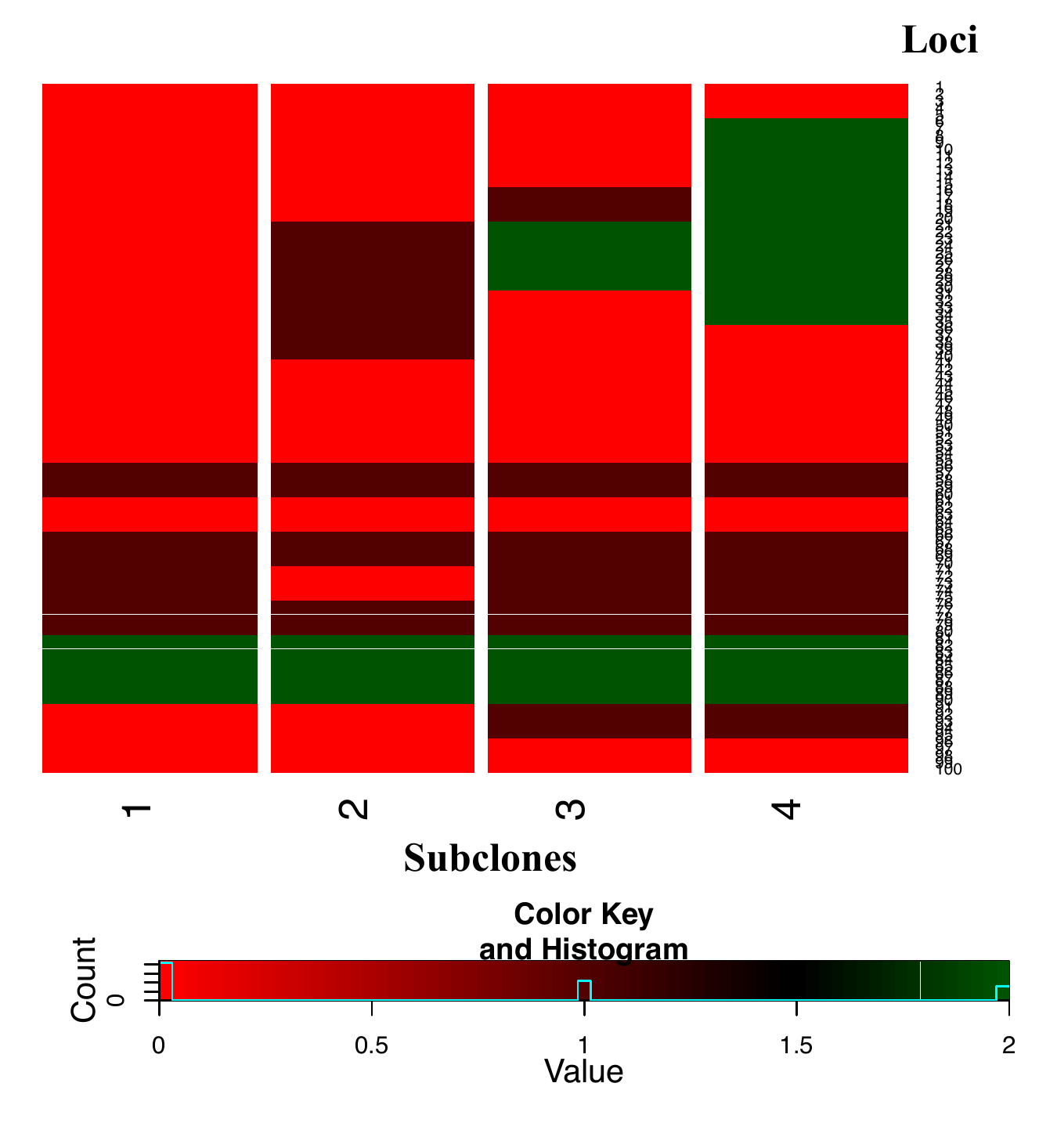} &
   \includegraphics[width=.32\textwidth]{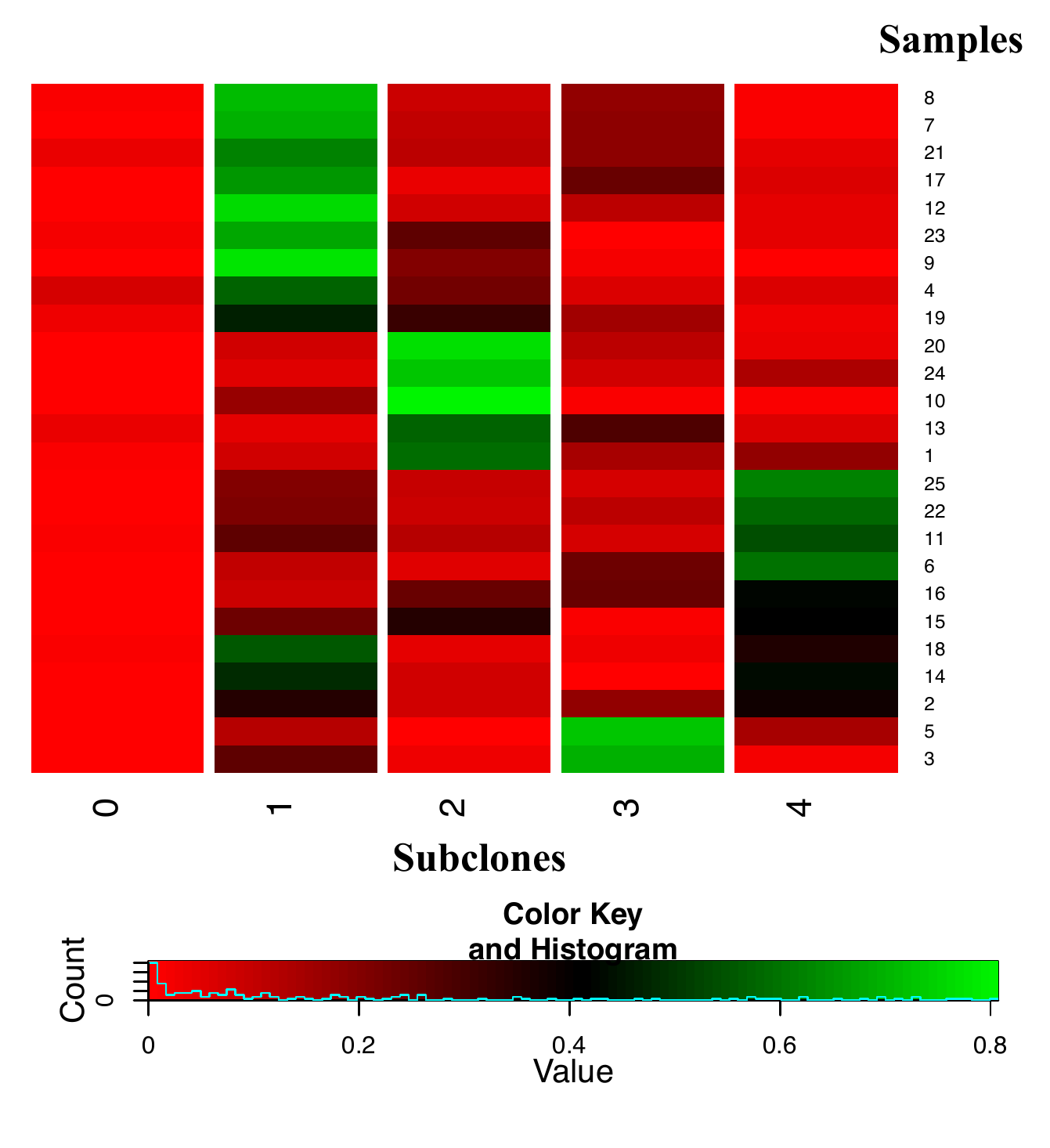} \\
  (a) $\bL^\true$ &
  (b) $\bZ^\true$ &
  (c) $\bw^\true$ 
\end{tabular}
 \end{center}
\caption{Simulation 2: simulation truth.}
\label{fig:Sim2_tr}
\end{figure}

%%%%%%%%%%%%%%%%%%%%%%%%%%%%%%%%%%%%%%%%%%%%%%%%%%%%%%%%%%%%%%%%%%%%%%%%%%%%%%%%%%%
We carried out a second simulation study with a more complicated subclonal structure.  We simulate read counts for a set of $S=100$ loci in $T=25$ hypothetical samples. 
In the simulation truth, we assume four latent subclones ($C^{\true}=4$) as well as a background
subclone ($c=0$) with all SNVs bearing variant sequences with two
copies. We use $Q=3$. The simulation truths, $\bL^\true$ and $\bZ^\true$ are shown in
Figure~\ref{fig:Sim2_tr}(a) and (b), respectively.
We generated $\phi^\true_t$ from $\Ga(600, 3)$ for $t=1, \ldots, 25$ and then
generated $\bw_{t}^\true$ as follows.  We let $\ba^\true=(13, 4, 2,
1)$ and for each $t$ randomly permuted $\ba^\true$.  Let
$\ba_\pi^\true$ denote a random permutation of $\ba^\true$.  We
generate $\bw^\true \sim \Dir(0.3, \ba_\pi^\true)$. That is, the
first parameter of the Dirichlet prior for the
$(C^\true+1)$-dimensional weight vector was $0.3$, and the remaining
parameters were a permutation of $\ba^\true$.  The weights $\bw^\true$
are shown in Figure~\ref{fig:Sim2_tr}(c).  The samples in the rows are rearranged for better display. 
From Figure~\ref{fig:Sim2_tr}(c), each sample has all the four
subclones with its own cellular fractions, resulting in large
heterogeneity within a sample.  In addition, the random permutation of
$\ba^\true$ induces heterogeneity among the samples.  We observe that
when the underlying subclonal structure is complicate and samples are
heterogeneous, larger sample size is needed. In particular, $T=25$
which is a large number compared to the typical sample size in 
real datasets is assumed for this simulation study. Using the assumed
$\bL^\true$, $\bZ^\true$ and $\bw^\true$ and letting $p_0^\true=0.05$,
we generate $N_{st} \sim \Poi(\phi_t^\true M_{st}^\true/2)$ and
$n_{st} \sim \Binom(N_{st}, p^\true_{st})$.  We fit the
proposed model as in the first simulation study.

%We carried out a second simulation study.  The simulation truth
%$\bL^\true$, $\bZ^\true$ and $\bw^\true$ is shown in
%Figure~\ref{fig:Sim2_tr}.  Again, we assumed $C^\true=4$. We simulated
%data and fit the proposed model as in the first simulation study.

%%%%%%%%%%%%%%%%%%%%%%%%
%%%%%%%%%%%%%%%%%%%%%%%%%

\begin{figure}[ht!]
  \begin{center}
\begin{tabular}{ccc}
   \includegraphics[width=.32\textwidth]{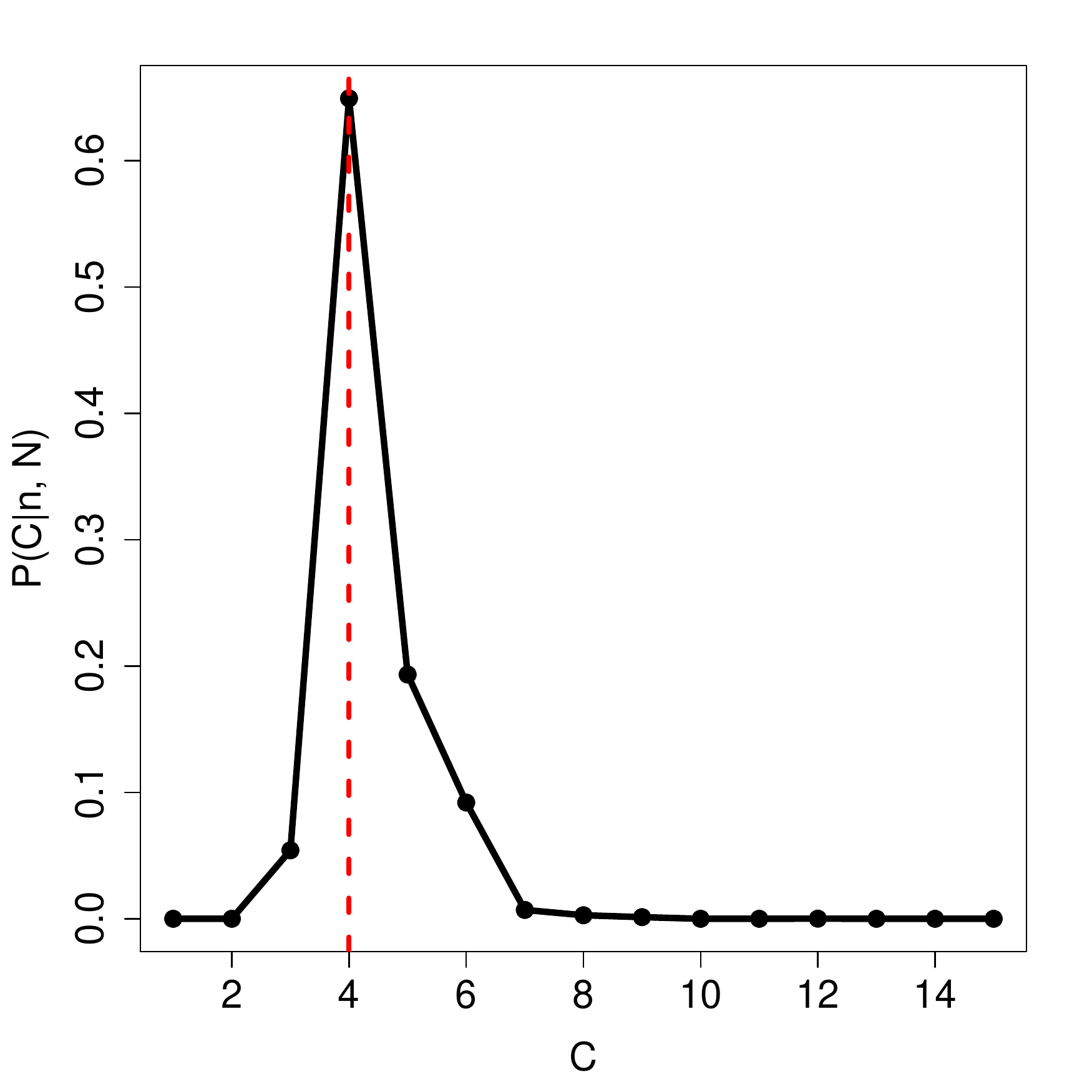} &
   \includegraphics[width=.32\textwidth]{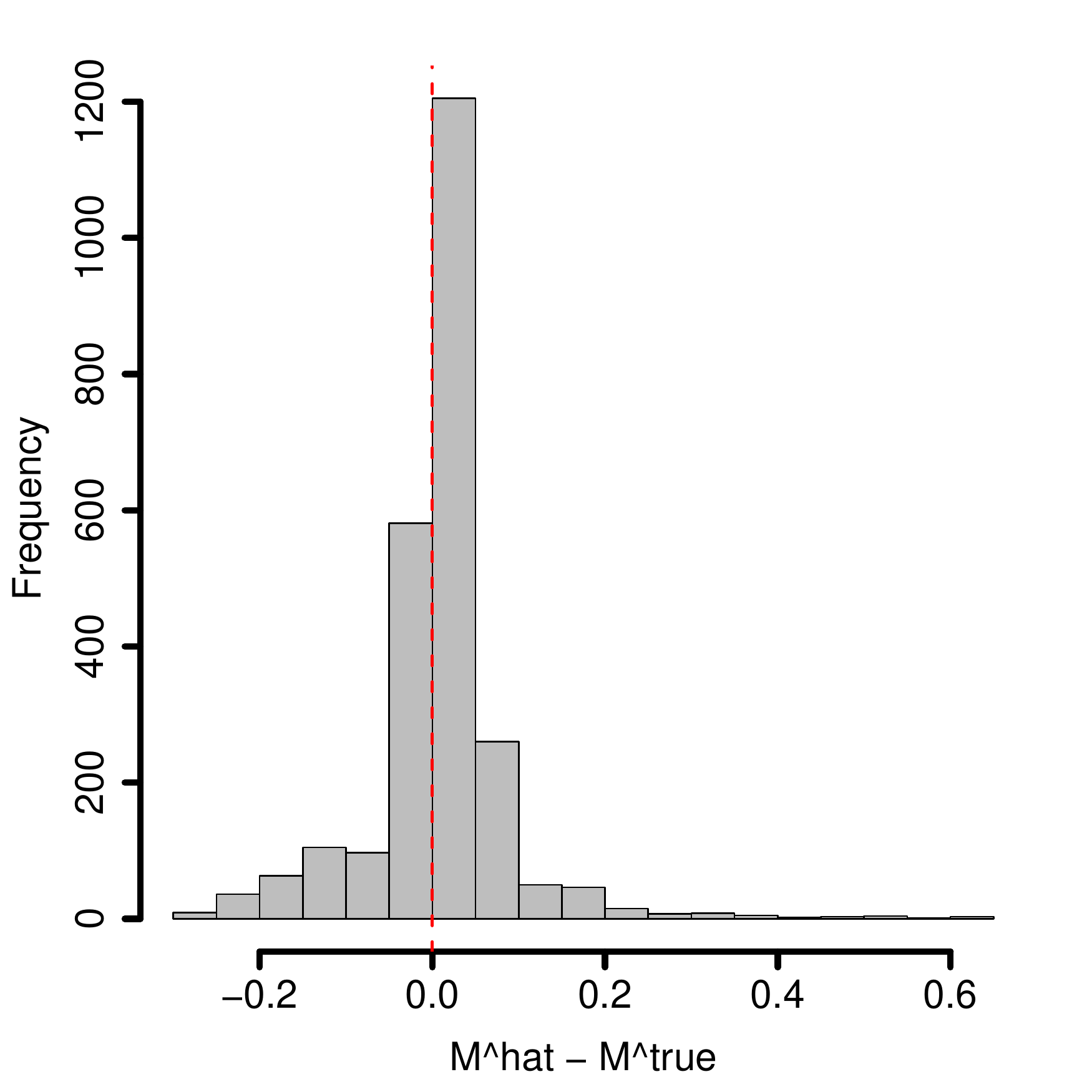} &
   \includegraphics[width=.32\textwidth]{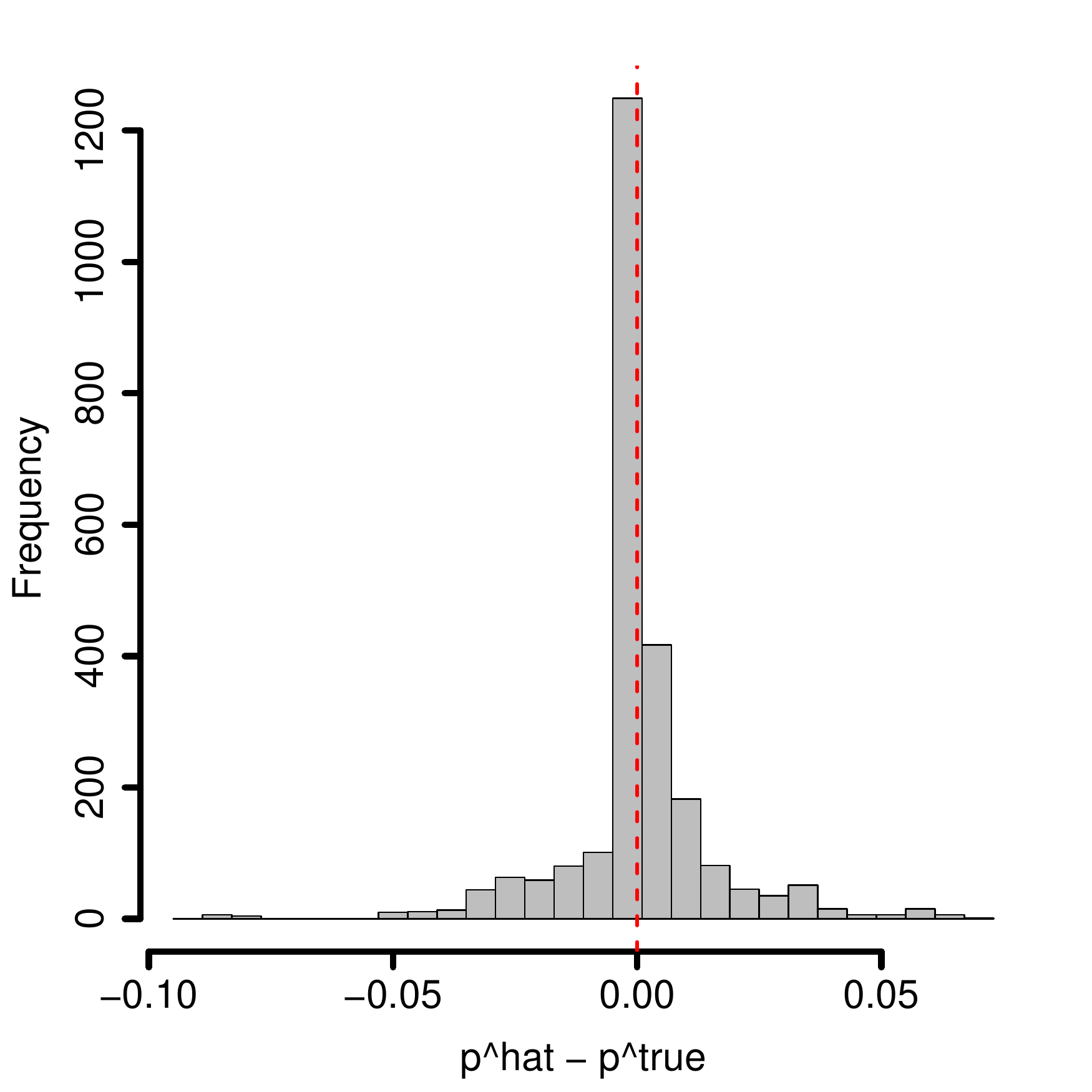} \\
  (a) $p(C \mid \bn,\bN)$ &
  (b) $\hat{M}_{st}- M^\true_{st}$ &
  (c) $\hat{p}_{st}- p^\true_{st}$ \\
   \includegraphics[width=.32\textwidth]{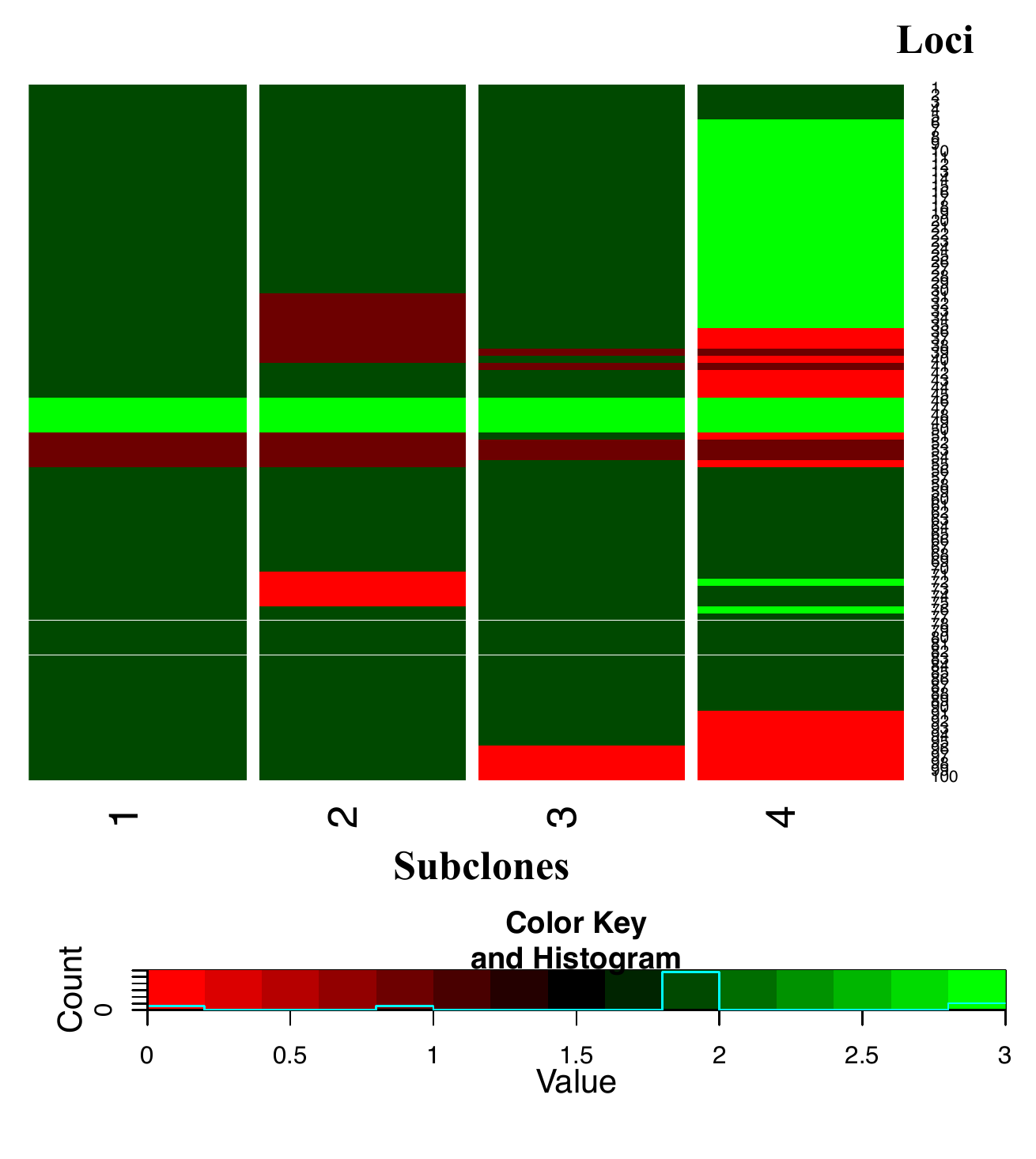} &
   \includegraphics[width=.32\textwidth]{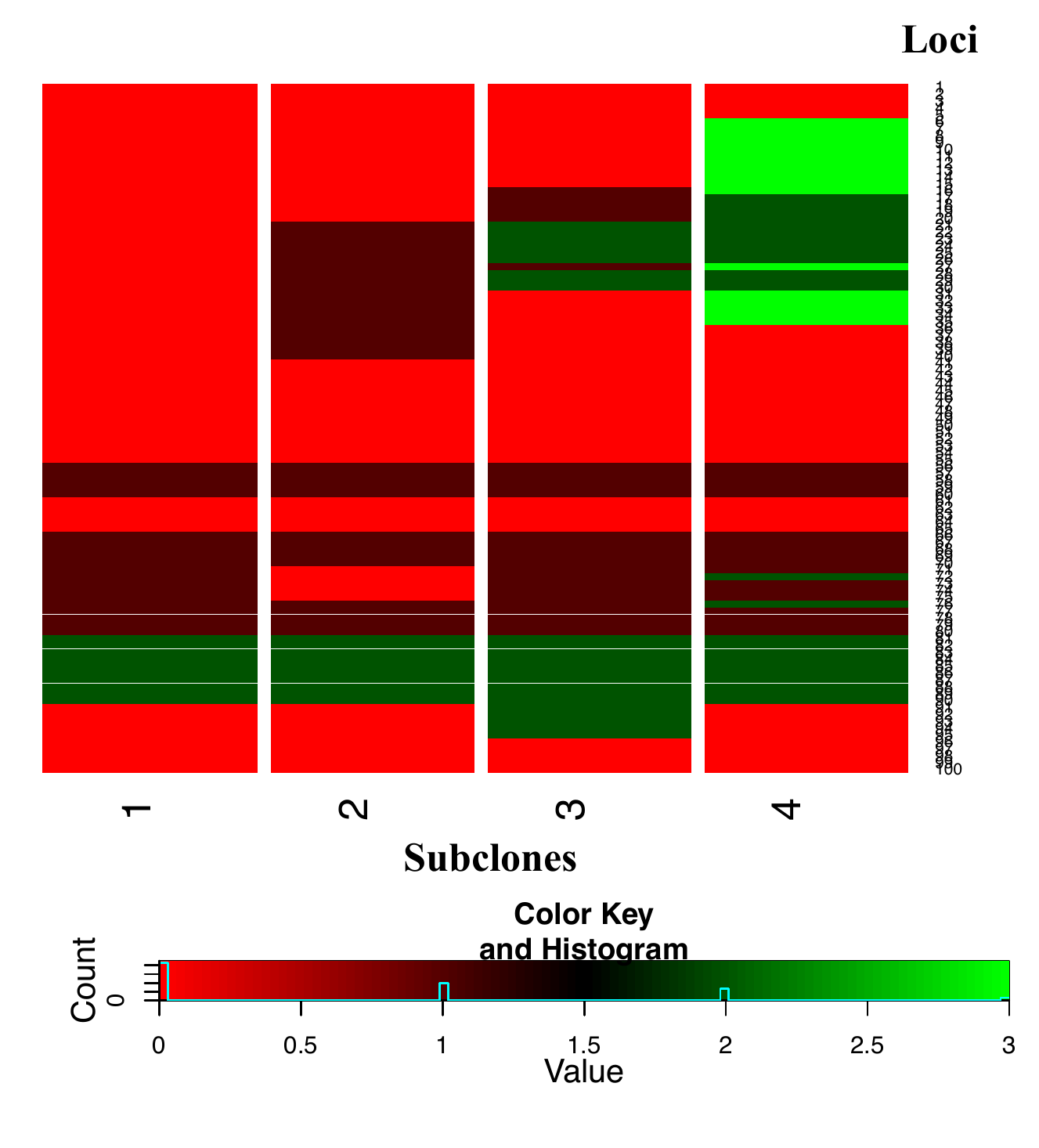} &
   \includegraphics[width=.32\textwidth]{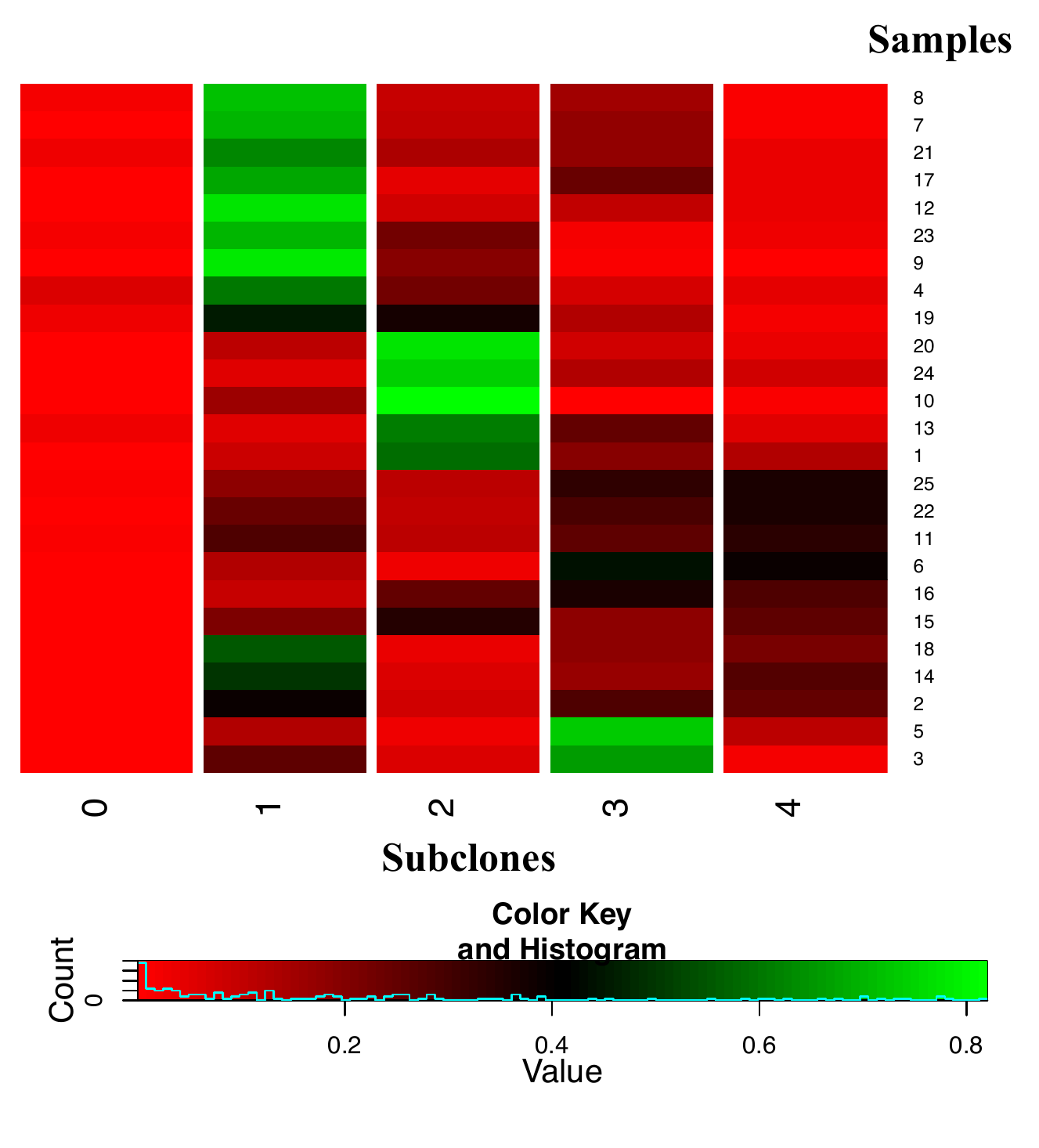} \\
  (d) $\bL^\star$ &
  (e) $\bZ^\star$ &
  (f) $\bw^\star$ \\
\end{tabular}
 \end{center}
\caption{Posterior inference for Simulation 2.}
\label{fig:Sim2_post}
\end{figure}

Figure~\ref{fig:Sim2_post}(a) reports  $p(C \mid \bn, \bN)$,
again marking $C^\true$ with a dashed vertical line. 
The posterior mode $\Cstar=4$ correctly recovers the truth.
Panels (d) through (f) summarize the posterior point estimates,
$\bL^\star$, $\bZ^\star$ and $\bw^\star$.  
% Compared to the simulation truth in Figure~\ref{fig:Sim2_tr}, 
 Posterior estimates accurately recover the simulation truth for
subclones 1 and 2,  for which the true proportions $w_{tc}^\true$ 
are large for many samples, as shown in
Figure~\ref{fig:Sim2_tr}(c).  
On the other hand, the posterior
estimate for subclones 3 and 4 shows discrepancies with the simulation truth.
In particular, we
observe that a group of loci that have $\ell^\true_{sc}=1$ in subclones 3
and 4 has $\ell^\star_{sc}=2$ for subclone 3 and
$\ell^\star_{sc}=0$ for subclone 4.  
We suspect that this  reflects the
small weights $w^\true_{tc}$ for $c=3,4$ for almost all samples, as seen in the
last two columns of Figure~\ref{fig:Sim2_tr}(c).  
 Notice also the bias in the corresponding estimates,
 $z^\star_{sc}$ and  $\bw^\star_{tc}$, $c=3,4$.   
% Similar to the previous simulation study,
Despite ambiguity about the true latent structure, 
we find a good fit to the data.
Conditional on $\Cs$, we computed $\hat{M}_{st}$ and $\hat{p}_{st}$
and compared to the true values.  Figure~\ref{fig:Sim2_post}(b) and
(c) show the summaries that indicate a good fit.

%%%%%%%%%%%%%%%%%%%%%%%%
%%%%%%%%%%%%%%%%%%%%%%%%%

%%%%%%%%%%%%%%%%%%%%%%%%
\begin{figure}[ht!]
  \begin{center}
\begin{tabular}{cc}
   \includegraphics[width=.32\textwidth]{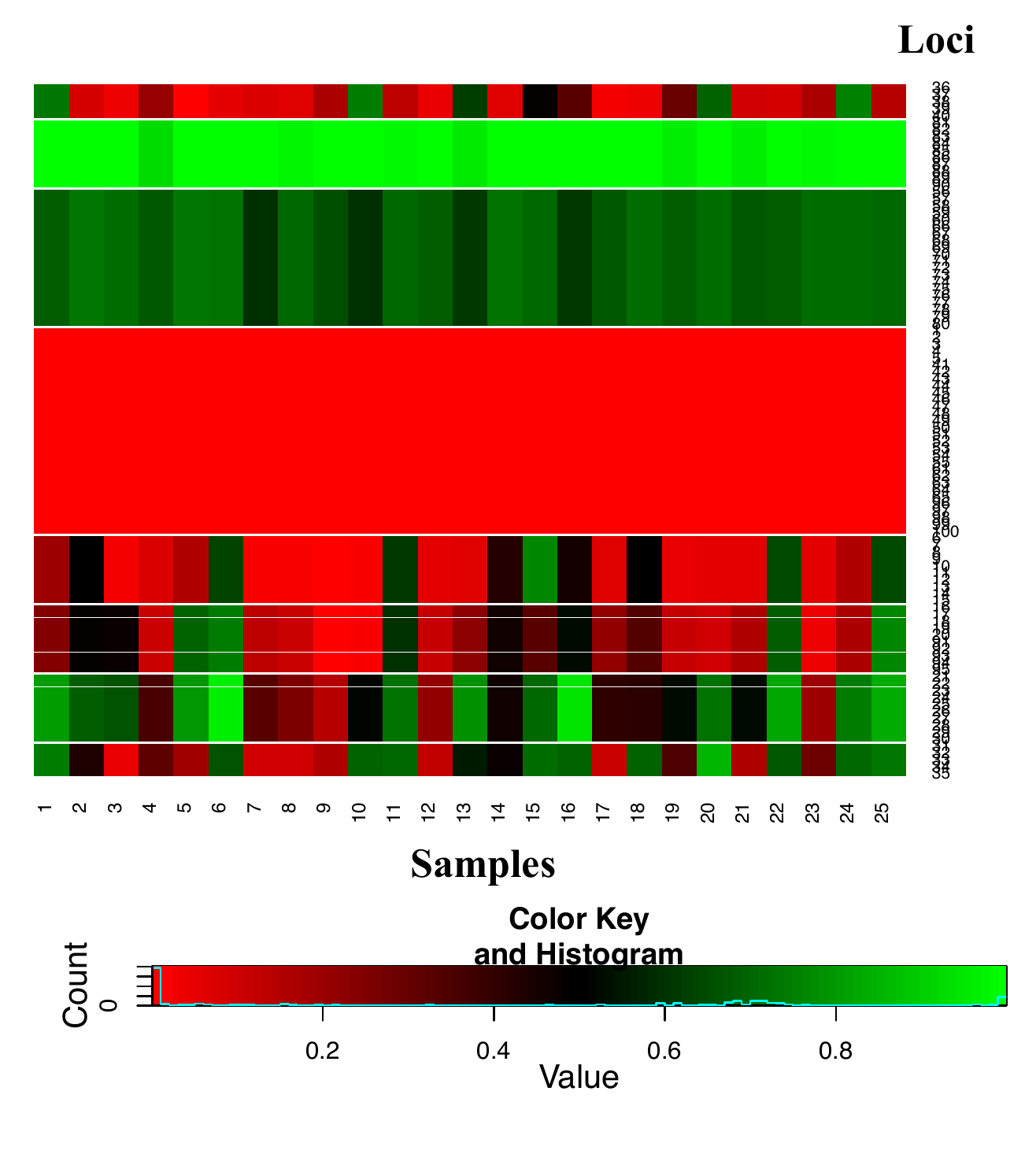} &
   \includegraphics[width=.32\textwidth]{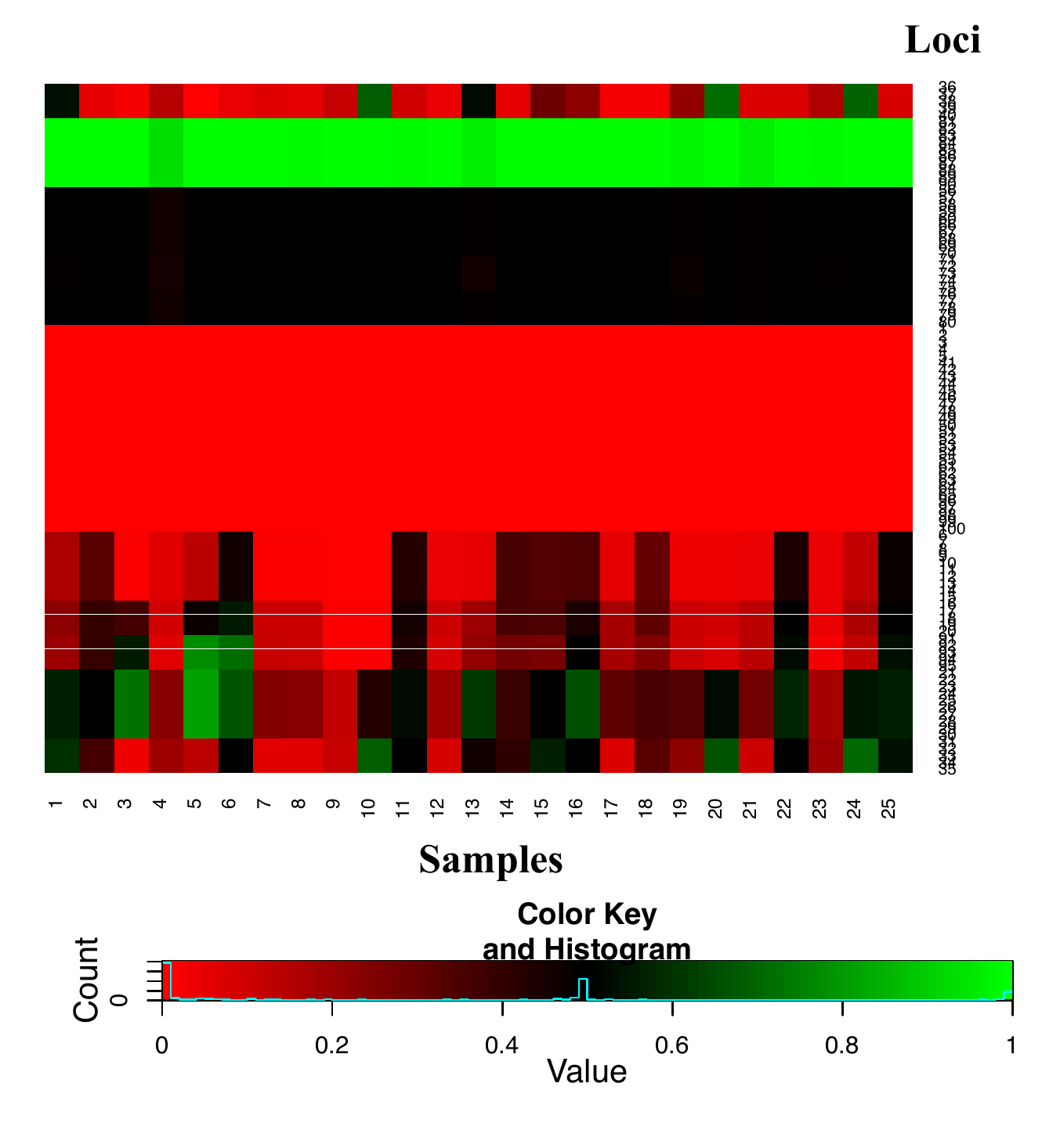} \\
  (a) Cellular prevalences &
  (b) $p^\true_{st}$ \\
\end{tabular}
 \end{center}
\caption{Heatmaps of estimated cellular prevalences from PyClone (a)
  and $p_{st}^\true$ (b) for Simulation 2.} 
\label{fig:Sim2_pyclone}
\end{figure}
%%%%%%%%%%%%%%%%%%%%%%%%%

For comparison, we again applied PyClone \citep{roth2014pyclone} to the same
simulated data. We used a similar setting for PyClone as in the
previous simulation.  Figure~\ref{fig:Sim2_pyclone}(a) shows 
 the estimated cellular prevalences. 
The reported clustering of loci (shown with by separations with white
horizontal lines) is reasonable. 
 Compare with the simulation truth  $p_{st}^\true$ in 
panel (b). 
The loci (rows) of the two heatmaps are re-arranged in the same order
for easy comparison. 
% By comparing the two heatmaps, the cellular
% prevalence estimates under PyClone are close to $p_{st}^\true$ and
% yields a reasonable estimates of  a clustering of the loci.  
Again,
PyClone does not attempt to reconstruct how subclones could
explain the observed data and does not provide inference on
the true subclonal structure in Figure~\ref{fig:Sim2_tr}.

%%%%%%%%%%%%%%%%%%%%%%%%%%%%%%%%%%%%%%%%%%%%%%%%%%%%%%%%%%%%%%%%%%%%%%%%%%%%%
\section{Lung Cancer Data}
\label{sec:lung_cancer}
%%%%%%%%%%%%%%%%%%%%%%%%%%%%%%%%%%%%%%%%%%%%%%%%%%%%%%%%%%%%%%%%%%%%%%%%%%%%%%%%%%%
We record whole-exome sequencing for four surgically dissected tumor
samples taken from the same patient with lung cancer.  We extracted
genomic DNA from each tissue and constructed an exome library from
these DNA using Agilent SureSelect capture probes. The exome library
was then sequenced in paired-end fashion on an Illumina HiSeq 2000
platform.  About 60 million reads - each 100 bases long - were
obtained.  Since the SureSelect exome was about 50 Mega bases, raw
(pre-mapping) coverage was about 120 fold.  We then mapped the reads
to the human genome (version HG19) \citep{HG_19} using BWA \citep{BWA}
and called variants using GATK \citep{GATK}.  Post-mapping, the mean
coverage of the samples was between 60 and 70 fold.

\begin{figure}[ht!]
  \begin{center}
\begin{tabular}{cc}
   \includegraphics[width=.4\textwidth]{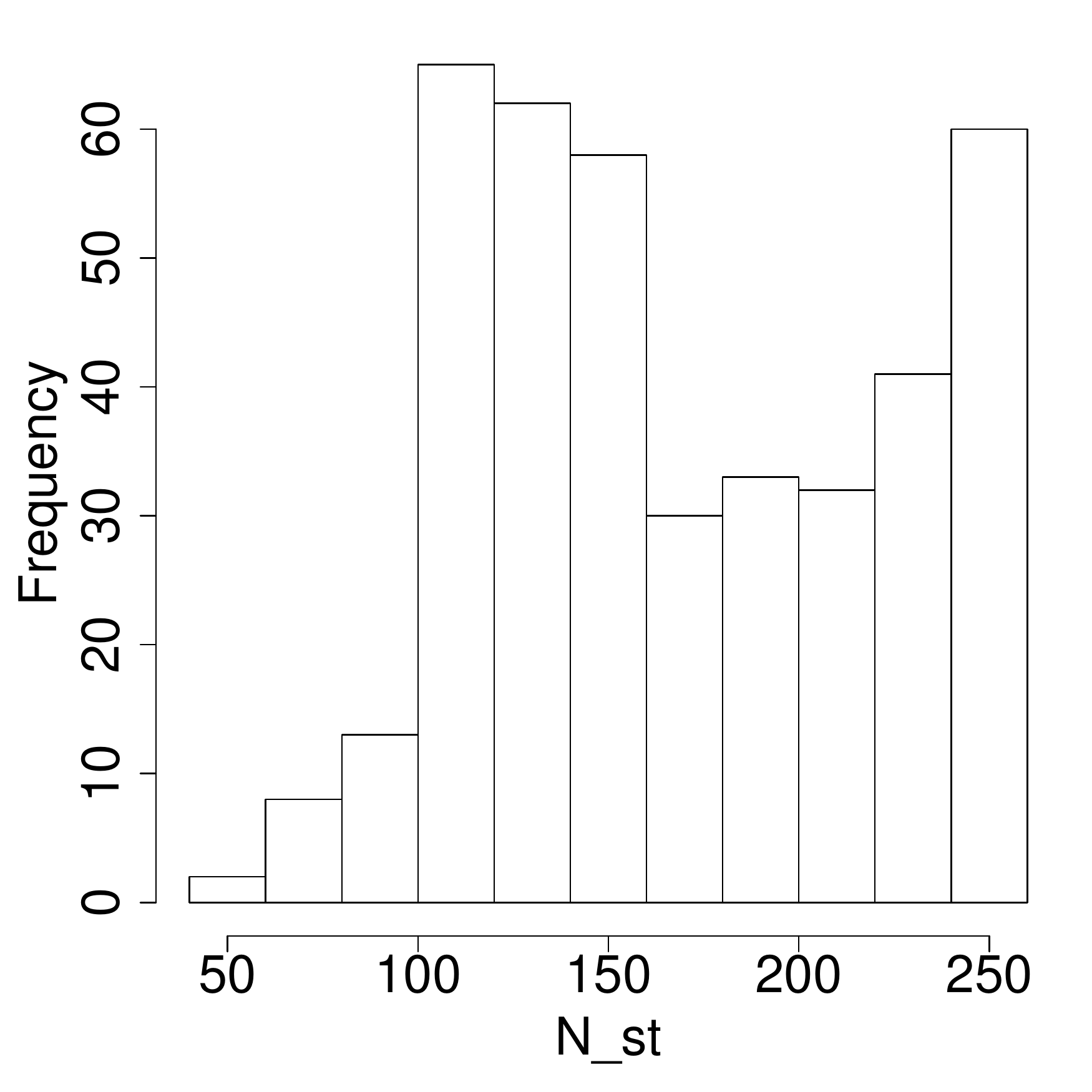} &
   \includegraphics[width=.4\textwidth]{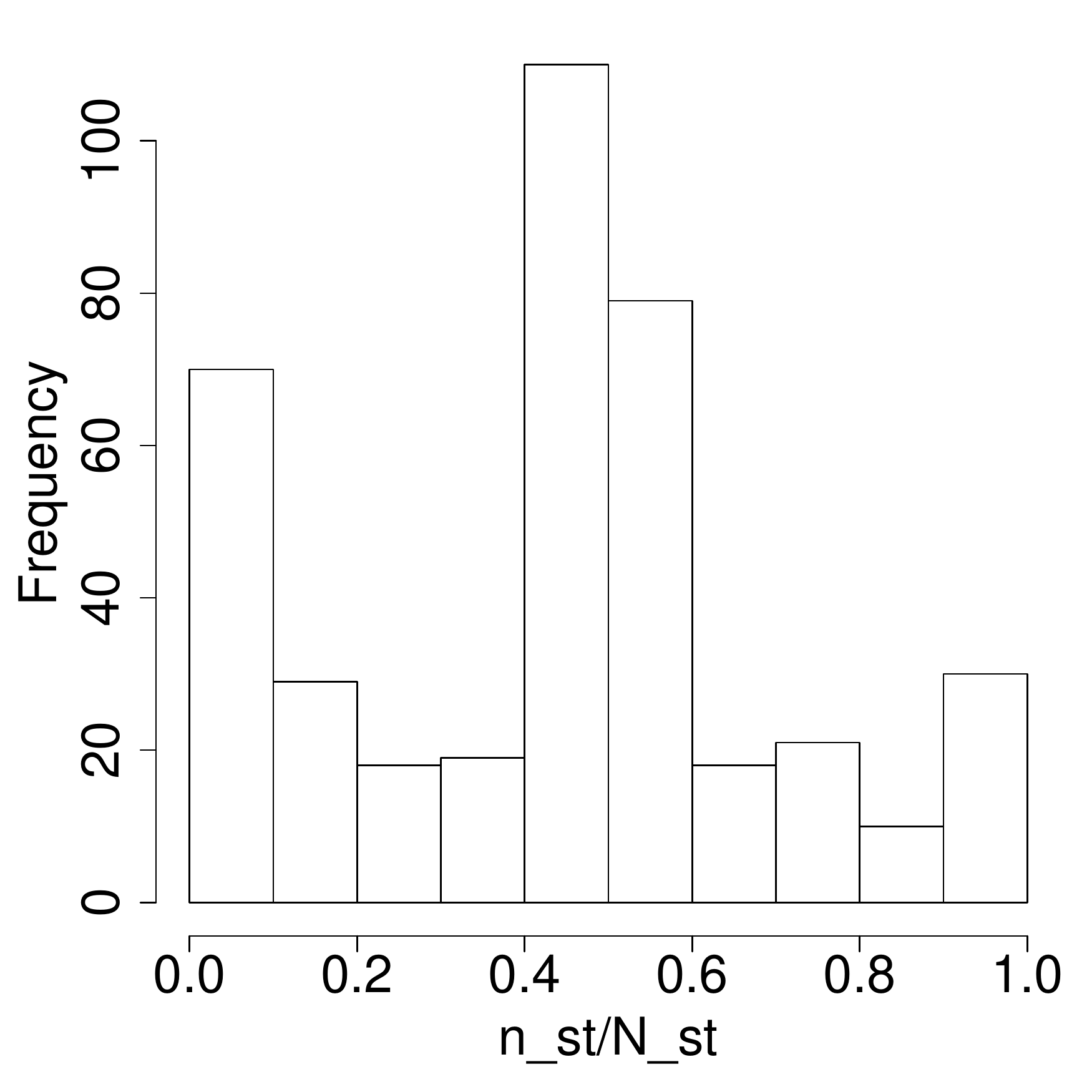} \\
  (a) Histogram of $N_{st}$ &
  (b) Histogram of $n_{st}/N_{st}$ \\
\end{tabular}
 \end{center}
\caption{Histograms of the Lung Cancer Dataset.}
\label{fig:Lung_data}
\end{figure}

A total of nearly 115,000 SNVs and small indels were called within the
exome coordinates. We restricted our attention to SNVs that (i) make a difference to the protein translated from the gene, and (ii) that exhibit significant coverage in all samples with $n_{st}/N_{st}$ not being too close to 0 or 1; and (iii) we used expert judgment to some more loci. 
% We restricted our attention to SNVs (i) that occur
%within genes that are annotated to be related to PDAC 
%\note{{\bf J:} i thought it's Lung cancer? PDAC probably stands for Pancreatic cancer?? --  \textbf{Yuan!}  Would you please modify this for the lung cancer?  I took this from the description on the pac cancer (we didn't have something like this for the lung cancer data.}
%in the KEGG
%pathways database \citep{KEGG}, (ii) that make a difference to the
%protein translated from the gene, and (iii) that exhibit significant
%coverage in all samples.   
% We obtained SNVs and filtered them based
% on criteria similar to the previous example, 
The described filter rules leave 
in the end $S=101$ SNVs for the four intra-tumor
samples. Figure~\ref{fig:Lung_data} shows the histograms of the total
number of reads and the empirical read ratios, $N_{st}$ and
$n_{st}/N_{st}$.

%%%%%%%%%%%%%%%%%%%%%%%%
%%%%%%%%%%%%%%%%%%%%%%%%%

%%%%%%%%%%%%%%%%%%%%%%%%
\begin{figure}[ht!]
  \begin{center}
\begin{tabular}{ccc}
   \includegraphics[width=.32\textwidth]{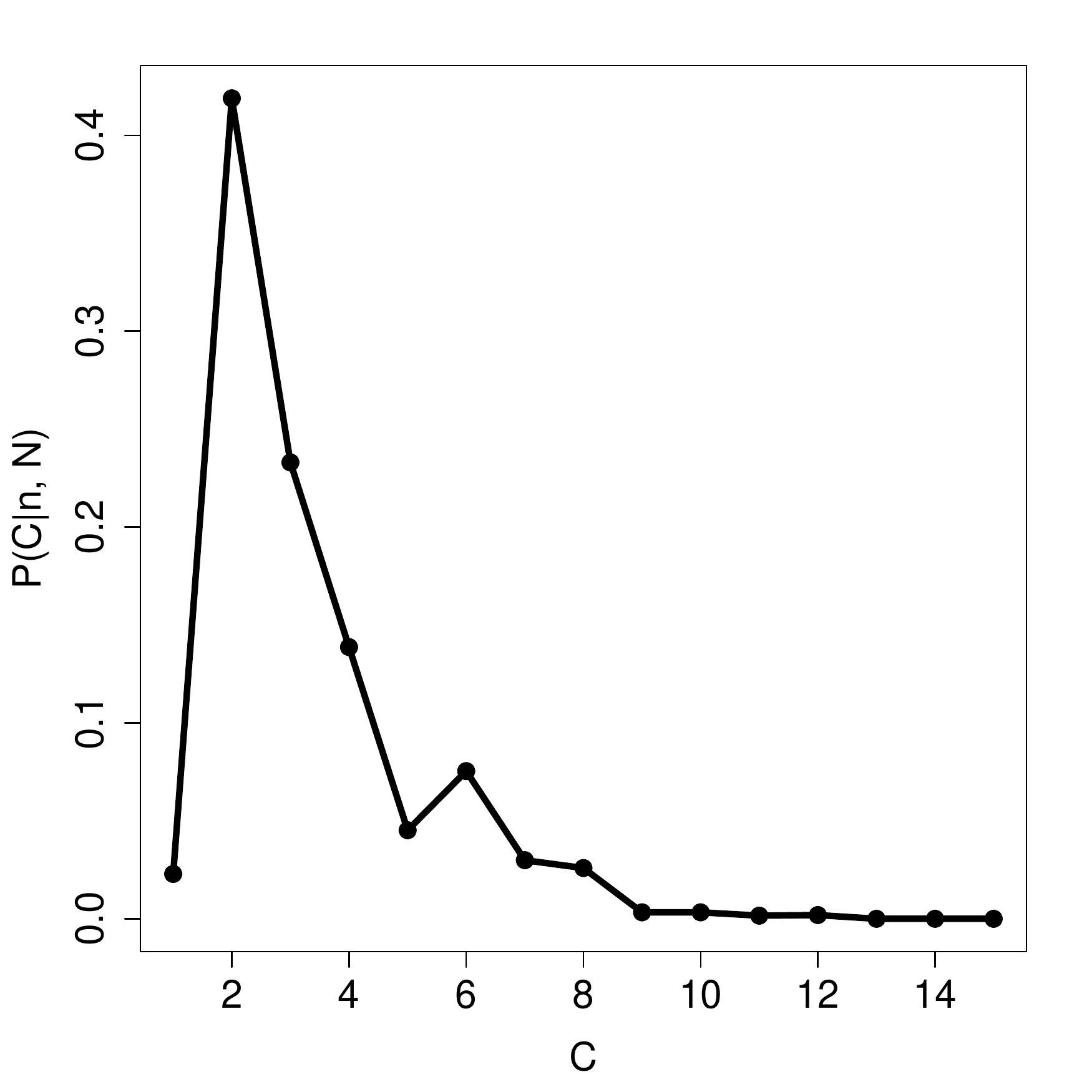} &
   \includegraphics[width=.32\textwidth]{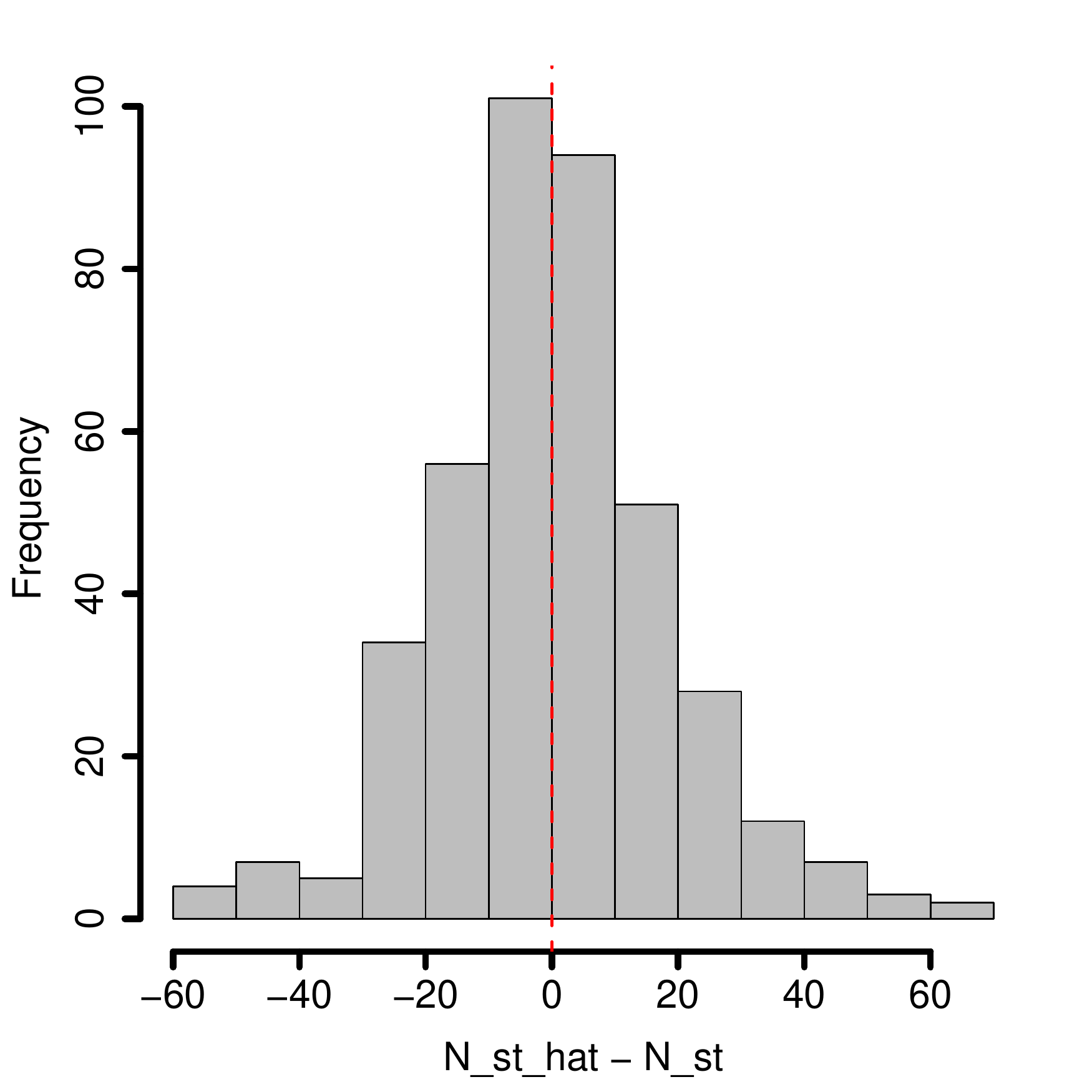} &
   \includegraphics[width=.32\textwidth]{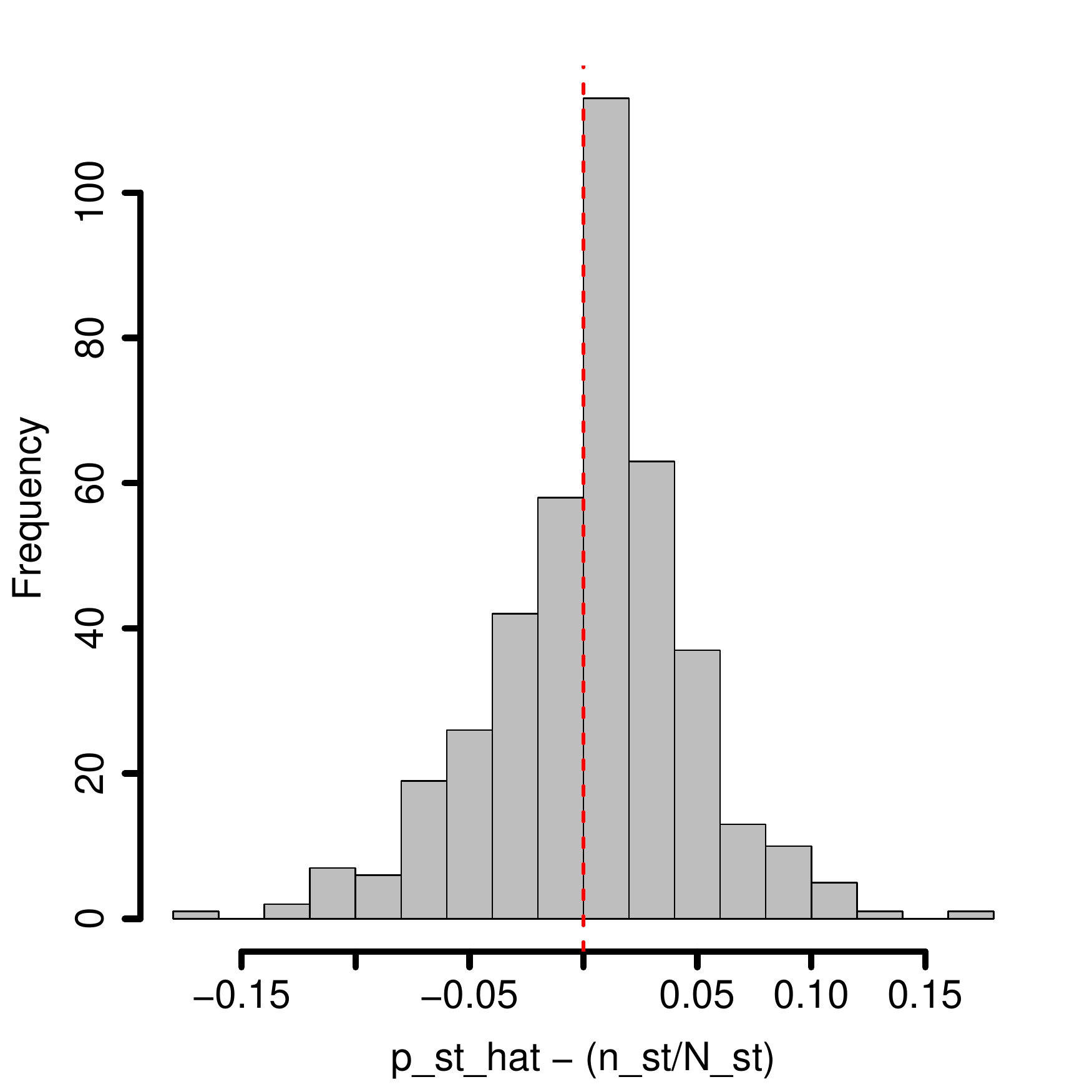} \\
  (a) $p(C \mid \bn, \bN)$ &
  (b) $\hat{N}_{st}- N_{st}$ &
  (c) $\hat{p}_{st}- (n_{st}/N_{st})$ \\
   \includegraphics[width=.32\textwidth]{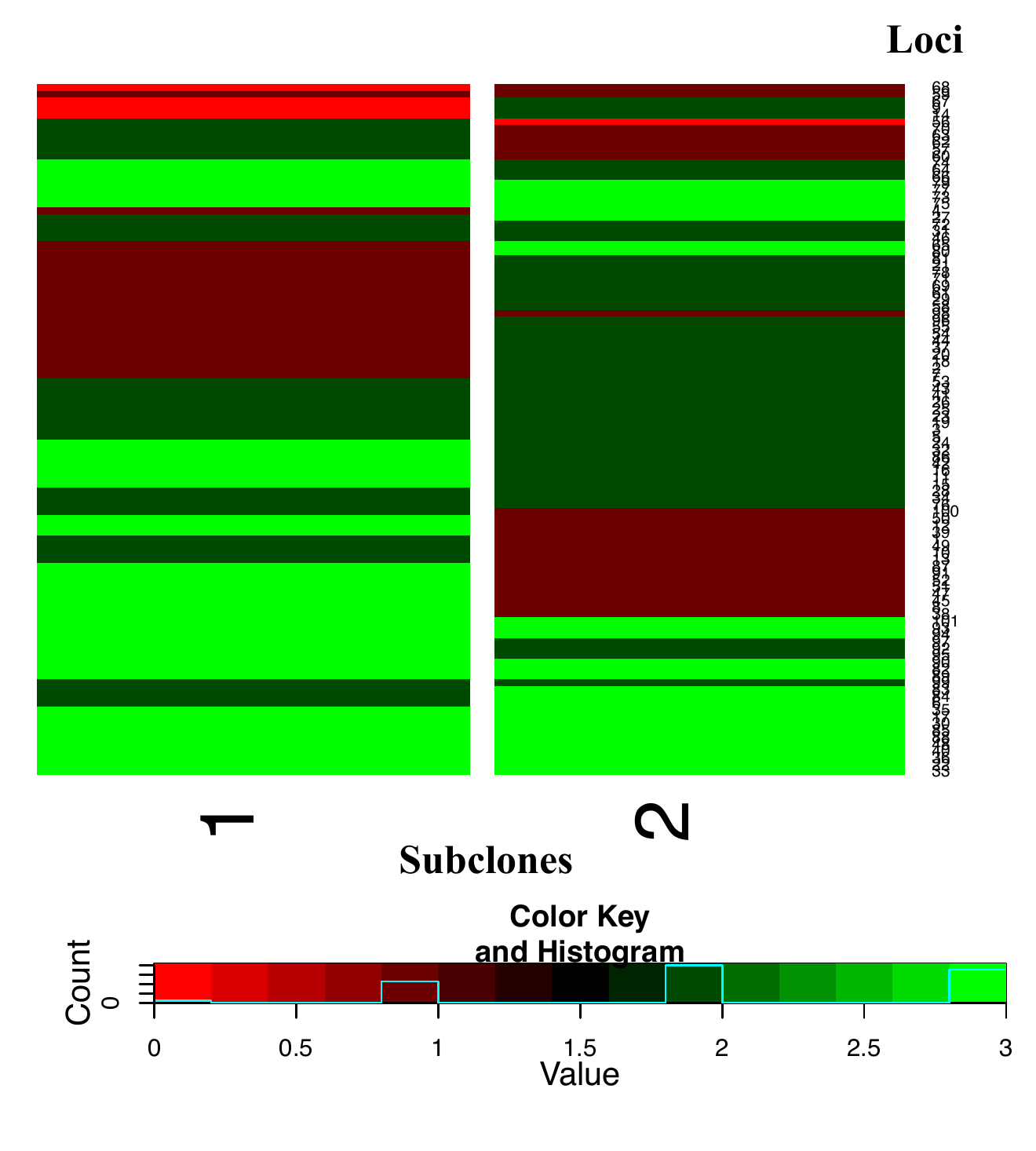} &
   \includegraphics[width=.32\textwidth]{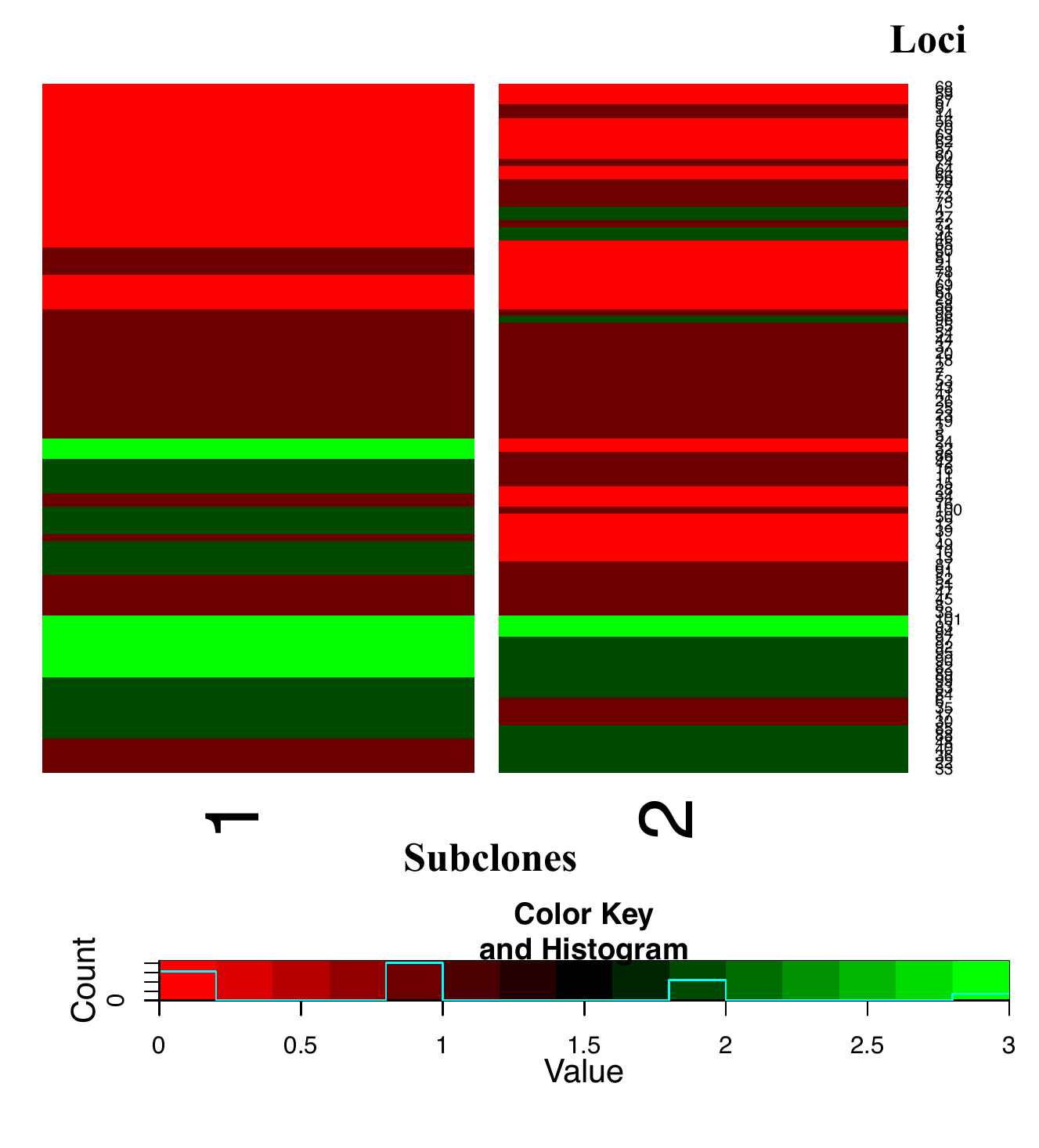} &
   \includegraphics[width=.32\textwidth]{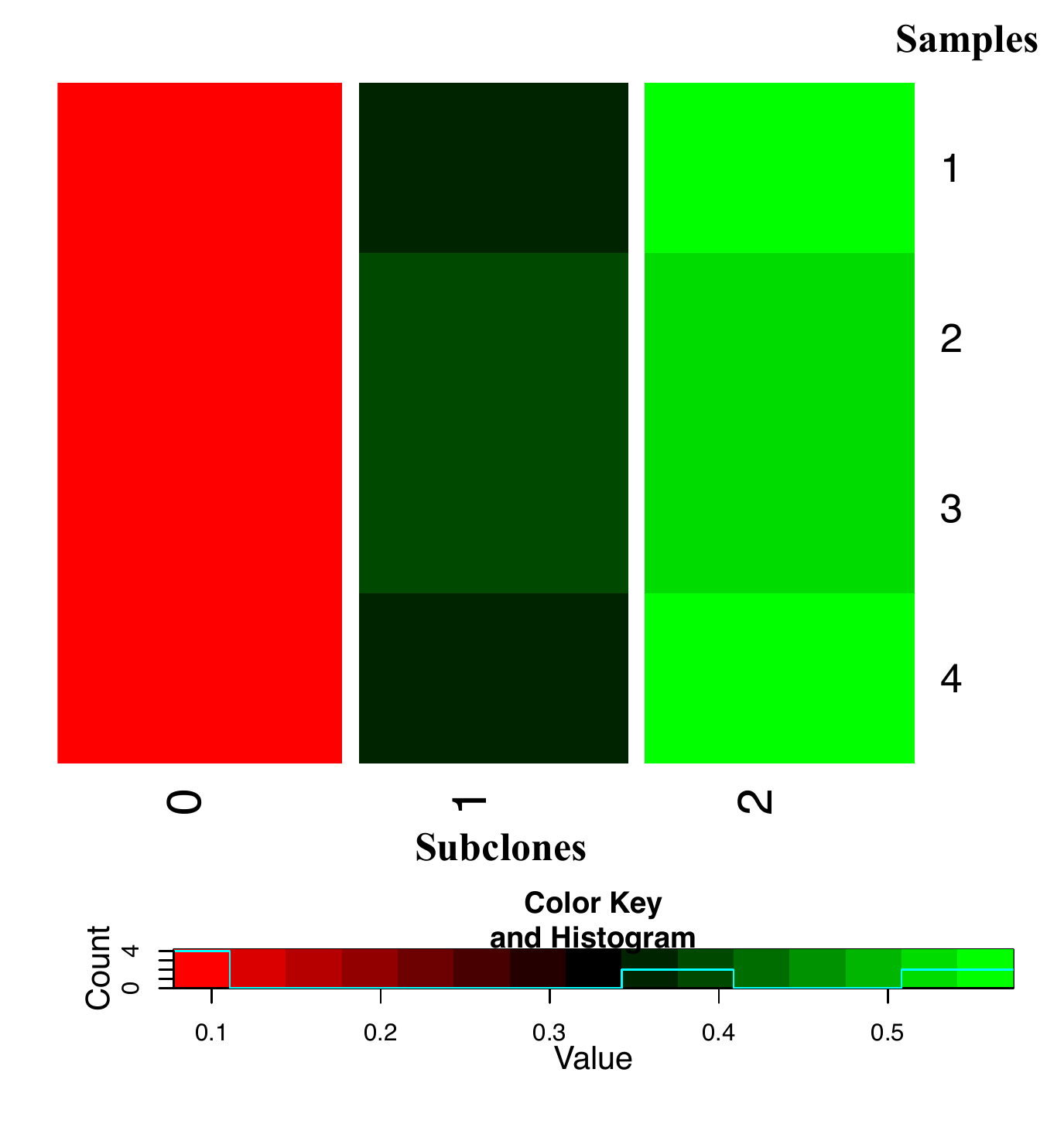} \\
  (d) $\bL^\star$ &
  (e) $\bZ^\star$ &
  (f) $\bw^\star$ \\
\end{tabular}
 \end{center}
\caption{Posterior inference for the Lung Cancer Dataset.}
\label{fig:Lung_post}
\end{figure}
%%%%%%%%%%%%%%%%%%%%%%%%%

We used hyperparameters similar to those in the simulation
studies.  Figure~\ref{fig:Lung_post} summarizes posterior
inference under the proposed model.  Panel (a) shows $\Cs=2$,  
i.e., two estimated subclones.     Using posterior samples with $C=\Cs$, we
computed $\hat{N}_{st}$ and $\hat{p}_{st}$ and compared them to the
observed data. The differences are centered at 0, 
implying a good fit to the data.
Conditional on $\Cs=2$, we found $\bL^\star$, $\bZ^\star$
and $\bw^\star$.  The loci in $\bL^\star$ and $\bZ^\star$ are
re-arranged in the same order for better illustration.  From
Figure~\ref{fig:Lung_data}(a) we notice that  many positions 
have large numbers of reads, over 200 reads. This 
is reflected in $\bL^\star$ which estimates three copies at many positions. 
The estimated weights $\bw^\star$ in Figure~\ref{fig:Lung_post}(f)
 show a great similarity across the four samples.  This lack of
heterogeneity across samples is not surprising. The four samples were
dissected from close-by spatial locations in the tumor.

%%%%%%%%%%%%%%%%%%%%%%%%
\begin{figure}[ht!]
  \begin{center}
\begin{tabular}{cc}
   \includegraphics[width=.4\textwidth]{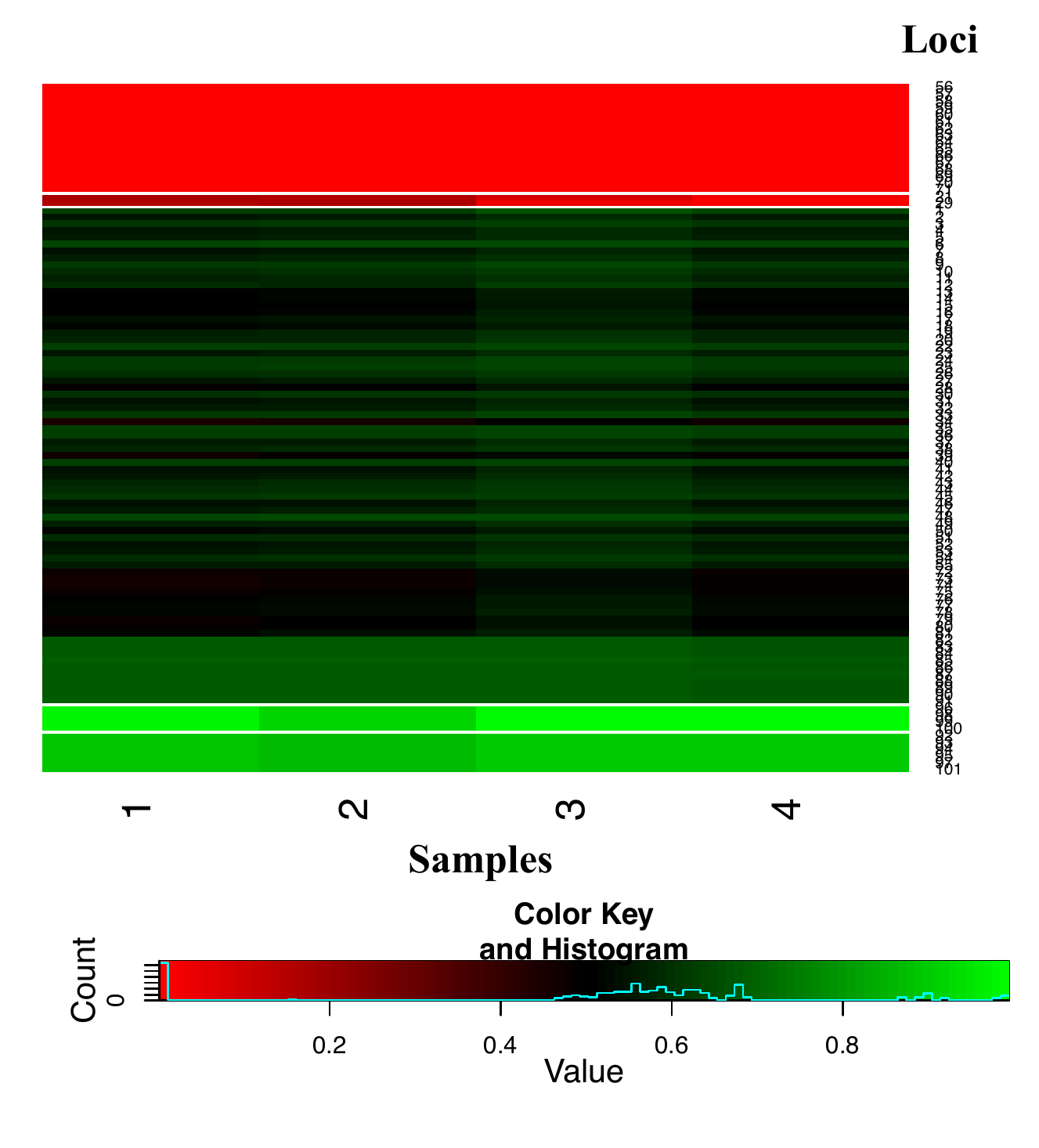} &
   \includegraphics[width=.4\textwidth]{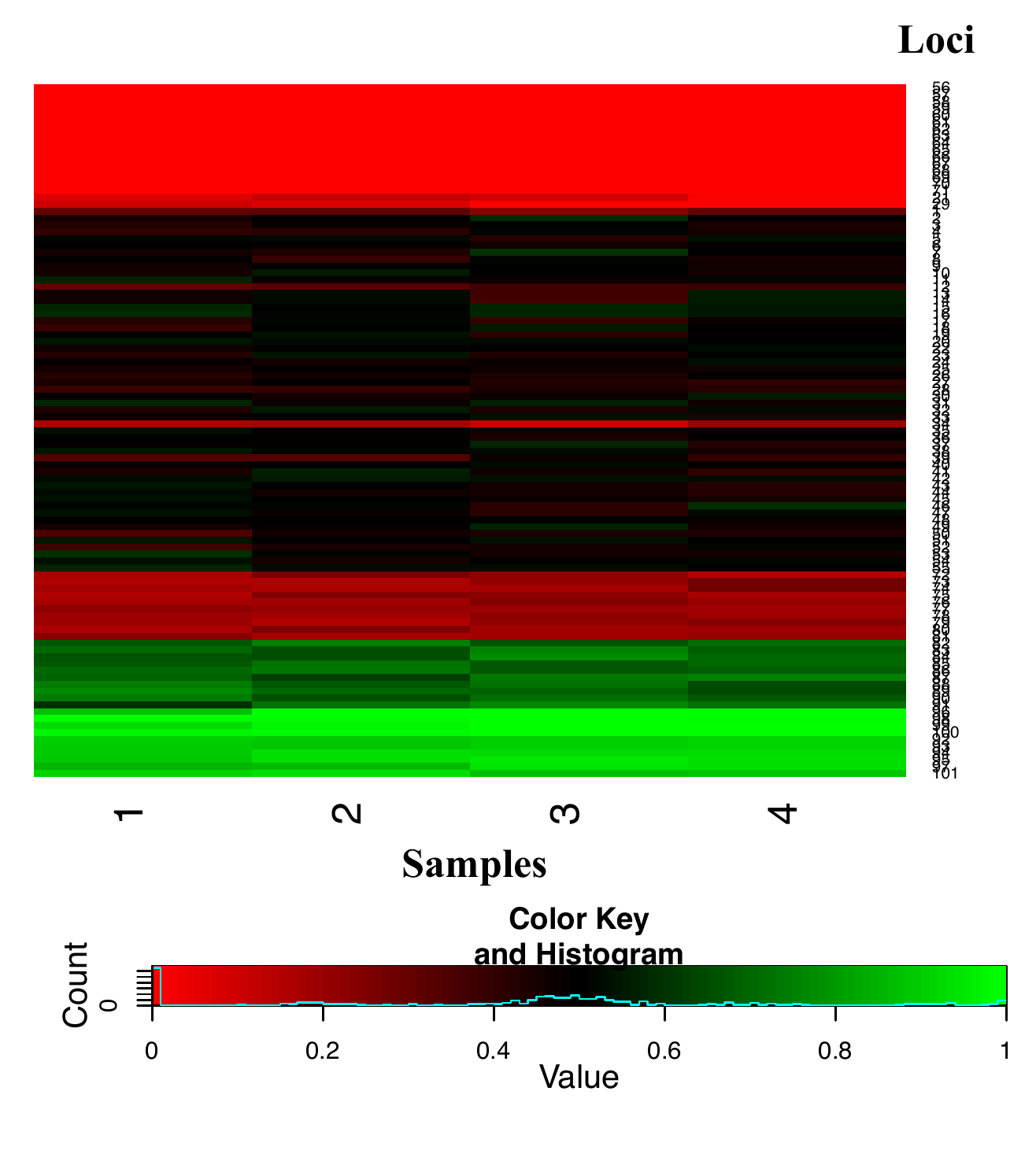} \\
  (a) Cellular prevalences &
  (b) $n_{st}/N_{st}$ \\
\end{tabular}
 \end{center}
\caption{Heatmaps of estimated cellular prevalences from PyClone (a) and $(n_{st}/N_{st})$ (b) for the Lung cancer dataset.}
\label{fig:Lung_pyclone}
\end{figure}
%%%%%%%%%%%%%%%%%%%%%%%%%

Again, for comparison  
% we used the same setting for the copy numbers as the previous
% simulation studies and
implemented PyClone \citep{roth2014pyclone} for the lung cancer
data. The  posterior estimates of prevalence and the estimated
clustering of the loci are shown in Figure~\ref{fig:Lung_pyclone}(a). The
clustering identified five clusters of the loci.  The mean prevalences within a
locus cluster are similar across samples, which is similar to
$\bw^\star$ in Figure~\ref{fig:Lung_post}(f). Panel (b)
of Figure~\ref{fig:Lung_pyclone} is  a heatmap of fractions
of reads  bearing mutation  for each locus and
sample. 
% From the
% comparison of the heatmap in panel (a) to that in panel (b), 
Again, PyClone
provides a reasonable estimate of a loci clustering based on the empirical
fractions, but does not provide an inference on subclonal populations.

%%%%%%%%%%%%%%%%%%%%%%%%%%%%%%%%%%%%%%%%%%%%%%%%%%%%%%%%%%%%%%%%%%%%%%%%%%%%%
\section{Conclusions}
\label{sec:Conclusion}
%%%%%%%%%%%%%%%%%%%%%%%%%%%%%%%%%%%%%%%%%%%%%%%%%%%%%%%%%%%%%%%%%%%%%%%%%%%%%%%%%%%
The proposed approach infers subclonal DNA  copy numbers,
 subclonal variant allele counts and cellular fractions   in a biological sample.  By
jointly modeling CNV and SNV,  we provide the desired description
of TH based on DNA variations in both, sequence and structure. Such
inference will significantly impact downstream treatment of individual
tumors, ultimately allowing personalized prognosis. For example, tumor with large proportions of cells bearing somatic mutations on
tumor suppressor genes should be treated differently from those that
have no or a small proportion of such cells. In addition, metastatic
or recurrent tumors may possess very different compositions of
cellular genomes and should be treated differently.  Inference on
TH can be exploited for improved treatment strategies for relapsed
cancer patients, and can spark significant improvement in cancer treatment in practice.

A number of extensions are possible for the present model. For
example, sometimes additional sources of information on CNVs such as a
SNP array may be available.  We then extend the proposed model to
incorporate this information into the modeling of $\bL$.  Another
meaningful extension is to cluster patients on
the basis of the imputed TH, that is, we link a random partition and a
feature allocation model. This extension may help clinicians assign
different treatment strategies, and be the basis of adaptive
clinical trial designs.

Inference for TH is a critical gap in the current literature. The ability to precisely break down a tumor into a set of subclones with distinct genetics would provide the opportunity for breakthroughs in cancer treatment by facilitating individualized treatment of the tumor that exploits TH. It would open the door for cocktail type of combinational treatments, with each treatment targeting a specific tumor subclone based on its genetic characteristics. We believe that the proposed model may provide a integrated view on subclones
to explain TH that remains a mystery to scientists so far.

%%%%%%%%%%%%%%%%%%%%%%%%%%%%%%%%%%%%%%%%%%%%%%%%%%%%%%%%%%%%%%%%%%%%%%%%%%%%%%
\section*{Acknowledgment}
Yuan Ji and Peter M\"{u}ller's research is partially supported by NIH R01 CA132897.

%%%%%%%%%%%%%%%%%%%%%%%%%%%%%%%%%%%%%%%%%%%%%%%%%%%%%%%%%%%%%%%%%%%%%%%%%%%%%%%%%%%

\bibliographystyle{natbib}
\bibliography{bibfile_intra}
\end{document}